\documentclass[aps,twocolumn,showpacs,preprintnumbers,amsmath,amssymb,superscriptaddress,floatfix,nofootinbib]{revtex4}
\usepackage{graphicx}
\usepackage{epsfig}
\usepackage{epstopdf}
\usepackage{hyperref}
\usepackage{amsmath}
\allowdisplaybreaks[1]
\usepackage{amsfonts}
\usepackage{amssymb}
\usepackage{multirow}
\usepackage{makecell}
\usepackage[dvipsnames]{xcolor}
\usepackage{ulem}
\usepackage{array}
\usepackage{lipsum}
\usepackage{subfigure}

\usepackage{soul}

\begin{document}

\title{Box Singularity Conditions in Box Diagrams of Decay Processes}

\author{Chao-Wei Shen} \email{shencw@hdu.edu.cn}
\affiliation{School of Science, Hangzhou Dianzi University, Hangzhou 310018, China}

\author{Ming-Yang Duan} \email{duanmingyang@ucas.ac.cn}
\affiliation{University of Chinese Academy of Sciences (UCAS), Beijing 100049, China}


\author{Jia-Jun Wu} \email{wujiajun@ucas.ac.cn}
\affiliation{University of Chinese Academy of Sciences (UCAS), Beijing 100049, China}
\affiliation{Southern Center for Nuclear-Science Theory (SCNT), Institute of Modern Physics, Chinese Academy of Sciences, Huizhou 516000, China}

\date{\today}

\begin{abstract}

Since 1959, singularities within single-loop diagrams have been studied. They are believed to significantly influence our understanding of experimental observables.
In this study, we explore the singularities that arise from box diagrams in the decay process, which can be classified into two distinct categories.
A comprehensive analysis of box singularities has been conducted, wherein we have derived and presented specific conditional formulae to ascertain the occurrence of singularities, along with the related physical scenarios.

\end{abstract}

\maketitle

\section{Introduction}
\label{sec:intro}

Quantum chromodynamics (QCD) is the fundamental theory of strong interaction. 
The basic fields of QCD are quark field, antiquark field and gluon field, and gluons are exchanged to carry the strong interaction. 
A prominent feature of QCD is color confinement, which states that only the color singlet states can be detected experimentally. 
Therefore, the colorless hadrons are the smallest visible states, and a better understanding of the hadron states, including both the internal structures and the spectrum, is one of the most valid ways to comprehend QCD and remains a significant challenge in particle physics. 
On the other hand, difficulties exist in the low energy region in QCD, which is around a few hundreds of MeV, since nonperturbative dynamics dominate here, and it is still a mystery. 
Different from the perturbation calculation, the loop diagrams that reflect nonperturbative properties may play an important role in the amplitude of a reaction. 
Thus, it is of great significance to investigate the loop diagrams.

Seeking for various hadrons from the experimental observations, such as decay widths and cross sections, is a crucial task in hadron physics.
As a general rule, the peak structure observed in the experimental invariant mass spectrum is recognized as a resonance. 
However, in reality, it may also be generated by some other mechanisms.
In Ref.~\cite{Guo:2017jvc}, the authors proved that there are two types of singularities in the three-point loop diagrams: two-body threshold cusps and triangle singularities (TS).
Both of them can be reflected in the invariant mass spectrum, with the latter corresponding to the peak structure where the TS occurs.
Actually, other kinds of loop diagrams can also exhibit similar or different singularities.
Therefore, it is of great significance to distinguish whether the peak structure observed in an experiment originates from a usual resonance or a pure kinematic singularity.

The singularity of hadron loop diagrams was first studied in Ref.~\cite{Landau:1959fi} in 1959.
In that paper, L. D. Landau investigated the essence of singularity in the case of single loops and typically provided the specific contributions of TS.
Nevertheless, at that time, it was impossible to find an actual process that met the rigorous kinematic conditions of TS.
Until around 2010, some processes that satisfy such kinematic conditions can be measured experimentally, 
and then the study of TS began to arouse great interest~\cite{Wu:2011yx, Aceti:2012dj, Wu:2012pg, Mikhasenko:2015oxp, Achasov:2015uua, Liu:2015taa,
Liu:2015fea, Guo:2015umn, Guo:2016bkl, Bayar:2016ftu, Wang:2016dtb, Xie:2016lvs, Roca:2017bvy,
Debastiani:2017dlz, Samart:2017scf, Sakai:2017hpg, Pavao:2017kcr, Xie:2017mbe, Bayar:2017svj,
Liang:2017ijf, Oset:2018zgc, Dai:2018hqb, Dai:2018rra, Liang:2019jtr, Nakamura:2019emd, Du:2019idk,
Liu:2019dqc, Jing:2019cbw, Sakai:2020ucu, Sakai:2020fjh, Molina:2020kyu, Shen:2020gpw, Huang:2020kxf}.
For example, in Refs.~\cite{Liu:2015fea,Guo:2015umn,Bayar:2016ftu}, the kinematic effect from the $\Lambda(1890) \chi_{c1} p$ triangle loop is suggested to be taken into consideration to figure out the nature of the observed $P_c(4450)$ by the LHCb collaboration in the $\Lambda_b^0 \to J/\psi K^- p$ reaction~\cite{LHCb:2015yax}.
Ref.~\cite{Guo:2019twa} makes a very comprehensive review of the threshold cusps and triangle singularities in hadronic reactions and the line shape of TS is proved to be similar to that of a normal resonance.
On the other hand, these mechanisms would cause an enhancement to other effects, such as the isospin breaking effect.
In Refs.~\cite{Wu:2011yx,Aceti:2012dj,Wu:2012pg, Achasov:2015uua, Du:2019idk}, the $K^* \bar{K} K$ triangle loop is found to produce a narrow peak in the $\pi \pi$ invariant mass spectrum of $\eta(1405/1475) \to \pi\pi\pi$, whose isospin violation is anomalously large, due to the mass difference between charged and neutral $K$.
Such isospin violating effects are usually very small in the order of $1\%$ compare to the isospin conservation process, however, in the region between $K^+ K^-$ and $K^0 \bar{K}^0$ thresholds, the isospin breaking receives a significant enhancement and is about 10 times the normal cases.
After taking such processes into consideration, the result turns out to be in accordance with the experimental data~\cite{BESIII:2012aa}.
In addition to TS, box singularity is also discussed in Ref.~\cite{Landau:1959fi}, but it is not very comprehensive.
In a word, the singularity from the hadron loop can have visible effects and is highly worthy of study.

In this paper, we are aiming to derive the condition of box singularities analytically.
This article is organized as follows. 
Sec.~\ref{sec:TS} contains the main rules of the divergence of the loop integral, and we apply them to the case of triangle singularities.
In Sec.~\ref{sec:diagram}, we demonstrate the derivation of two different types of box loop diagrams followed by a discussion.
The conditions for singularities are detailed studied in Sec.~\ref{sec:diagramA} and Sec.~\ref{sec:diagramB}, and the relevant physical properties are also discussed. 
The specific condition formulae of the masses in the diagram for determining whether singularity can occur is derived and given in Sec.~\ref{sec:formula}.
At last, a brief summary of this work is presented in Sec.~\ref{sec:sum}.
Various technicalities, such as the detailed derivation, are provided in the appendices.

\section{The loop integral and triangle singularity} 
\label{sec:TS}

In loop calculations, in addition to the usual ultraviolet and infrared divergence, some singularities can lead to a new divergence.
This divergence will result in a peak-like structure in the invariant mass spectra of certain final states, which can be mixed with the peaks generated by resonances.
Since this singularity position is independent of the interaction term and only depends on the kinematic effect of all involved particles in the loop, it is useful to extract the position of such singularity in phase space.

\begin{figure}[htbp]
\centering
\includegraphics[width=0.9\linewidth]{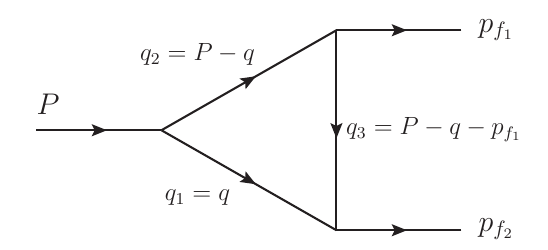}
\caption{The triangle loop diagram. \label{fg:2loop}}
\end{figure}

We start with the simplest single loop diagram of the $i \to f_1+f_2$ decay containing three intermediate particles to explore the integral structure that leads to singularity.
Such a triangle loop diagram is shown in Fig.~\ref{fg:2loop}, where the convention of notation and numeration are given.
The corresponding scalar loop integral writes: 
\begin{align}
     I_2 \equiv &  \frac{1}{(q^2-m_1^2+i\varepsilon)[(P-q)^2-m_2^2+i\varepsilon]} \nonumber \\
     & \times \frac1{\left[(P-q-p_{f_1})^2-m_{3}^2+i\varepsilon\right]}, \nonumber \\
    {\cal I}_2 \equiv &\int {\rm d}^4 q I_2, 
     \label{eq:loop2}
\end{align}
where $m_i (i=1,2,3)$ is the mass of the $i$-th intermediate particle.
Please note that small quantities such as $\varepsilon$ are taken to be $0^+$ unless otherwise specified.
It should be noted that the interaction vertex functions in the loop integration hereinafter are neglected, as they do not affect the existence of the singularity we are concerned about, although such vertexes will affect the strength of singularity.
Furthermore, such a triangle loop diagram is symmetric, telling that changing the direction of $q_3$ is equivalent.

In the rest frame of the initial particle, i.e. $P=(M,\vec{0})$, there are a total of six poles for the integral of $q^0$ in Eq.(\ref{eq:loop2}).
Three of them, $-\omega_1+i\varepsilon$, $M-\omega_2+i\varepsilon$ and $E_{f_2}-\omega_3+i\varepsilon$, are located above the real axis of the complex energy plane, while the other three, $\omega_1-i\varepsilon$, $M+\omega_2-i\varepsilon$, and $E_{f_2}+\omega_3-i\varepsilon$ are located below the real axis. 
Here $\omega_i = \sqrt{\vec{q}_i^2+m_i^2}$ is the on-shell energy of the $i$-th intermediate particle in the loop, which is a function of $\vec{q}_i$, and $E_{f_i}$ is the energy of final particle $f_i$.
Then we can obtain the integral of $q^0$ by calculating the residues of the poles as follows,
\begin{align}
    {\cal I}_2 =& \int {\rm d}^3 \vec{q} \ [  {\rm Res}(I_2, \omega_1-i\varepsilon) + {\rm Res}(I_2, M+\omega_2-i\varepsilon) \nonumber \\ 
    & + {\rm Res}(I_2, E_{f_2}+\omega_3-i\varepsilon) ],
\label{eq:loop2q0}
\end{align}
where ${\rm Res}(I, Z)$ represents the residue of the function $I$ at pole $Z$.

The integral variable can be decomposed as ${\rm d}^3 \vec{q} \equiv q^2 {\rm d}q {\rm d}\cos\theta {\rm d}\phi$.
We define $\vec{p}_{f_2} = (0,0,p_{f_2})$ with ($p_{f_2}\ge0$), so only two variables are left, i.e. ${\rm d}^3 \vec{q} = 2\pi q^2 {\rm d}q {\rm d}\cos\theta$.
The integral interval of $q$ is $[0,\infty)$ and that of $\cos\theta$ is $[-1,1]$.
Then we have $\omega_{1,2}(q)=\sqrt{q^2+m_{1,2}^2}$, $\omega_3(q,\theta)=\sqrt{p_{f_2}^2+q^2+m_3^2-2p_{f_2}q \cos\theta}$.
Besides, the singularity from the poles of these two variables in all residues is first-order, which is determined by the propagator of intermediate particles.
For such first-order poles, there are only two possible divergences after integration.
In the first case, the pole position is exactly on the boundary of the integration interval, such as $\int_{0}^{\text any} x^{-1} dx \to \infty$.
The second scenario is that the integral path is pinched between two poles, where the real parts of these two poles are the same, with one on the upper edge of the real axis and the other on the lower edge.

For the integral ${\cal I}_2$, it can be found that singularity only occurs at the residue of pole $\omega_1-i\varepsilon$.
Then ${\cal I}_2$ is written as:
\begin{align}
    {\cal I}_2=& \int {\rm d}^3 \vec{q} \Big[ \frac{f(\vec{q})}{(M-\omega_1(q)-\omega_2(q)+i\varepsilon) } 
    \nonumber \\
    & \times \frac{1}{(E_{f_2}-\omega_1(q)-\omega_3(q,\,\theta)+i\varepsilon)} + h(\vec{q}) \Big],
\label{eq:loop2final}
\end{align}
where $f(\vec{q})$ and $h(\vec{q})$ are general polynomials and do not produce singularity.
They do not affect the singularity position and divergence, as they become a constant when the integral variables approach the singularity position.
Therefore, we can just focus on the two denominator terms.
The first term $1/(M-\omega_1(q)-\omega_2(q)+i\varepsilon)$ has a pole at $q^+_{\rm on}= q_{\rm on}+i\varepsilon$, where $q_{\rm on}=\sqrt{\lambda(M^2,m_1^2,m_2^2)}/(2M)$ is the on-shell momentum of particles $1$ and $2$ in the rest frame of the initial particle and $\lambda(\alpha,\beta,\gamma)=\alpha^2+\beta^2+\gamma^2-2\alpha\beta-2\alpha\gamma-2\beta\gamma$ is the K\"all$\rm \acute{e}$n function.
To make the integral path pinched between two poles, the other term $1/(E_{f_2}-\omega_1(q)-\omega_3(q,\theta)+i\varepsilon)$ should have a pole $q^-_{\rm on}=q_{\rm on}-i\varepsilon^\prime$ after integrating $\cos\theta$.
It, which is a function of $q$ and $\cos\theta$, can be written as:
\begin{align}
  & \frac1{E_{f_2}-\omega_1(q)-\omega_3(q,\,\theta)+i\varepsilon}  \nonumber \\
  =& \frac1{E_{f_2}-\sqrt{m^2_1+q^2} - \sqrt{p_{f_2}^2+q^2+m_3^2-2p_{f_2}q \cos\theta}+i\varepsilon}.
\label{eq:second}
\end{align}
It can be seen that it only contains the first-order pole of $\cos\theta$.
To generate singularity, $\cos\theta$ of the pole position should be located at the boundary of the integration interval, which means $\cos\theta=-1$ or 1.
Therefore, the denominator of Eq.(\ref{eq:second}) has to be zero when $q=q^-_{\rm on}$ and $\cos\theta=-1$ or $1$, then we have:
\begin{align}
E_{f_2}-\omega_1(q_{\rm on})- \omega_3(q_{\rm on},\,\pi\,{\rm or}\,0)= 0, \\
i\varepsilon+i\varepsilon^\prime \left( \frac{q_{\rm on}}{\omega_1(q_{\rm on})} - \frac{\mp p_{f_2}-q_{\rm on}}{\omega_3(q_{\rm on},\,\pi\,{\rm or}\,0)} \right) = 0.
\label{eq:imaginary}
\end{align}
%
To make Eq.(\ref{eq:imaginary}) hold, the value of the term in the bracket needs to be negative, which means that the value of $\cos\theta$ can only be taken as $1$, i.e. $\theta = 0$.
Therefore, the singularity conditions of the triangle loop diagram are: 
\begin{align}
E_{f_2} &= \omega_1(q_{\rm on})+ \omega_3(q_{\rm on},\,0), \nonumber \\
v_1 &= \frac{q_{\rm on}}{\omega_1(q_{\rm on})} < \frac{p_{f_2}-q_{\rm on}}{\omega_3(q_{\rm on},\,0)} = v_3.
  \label{eq:2loopcondition}
\end{align}

From the physical perspective, $\cos\theta=1$ means that all involved particles are moving on the same line, and $v_1< v_3$ indicates that the intermediate particle 3 and particle 1 are moving in the same direction, and particle 3 moves faster than particle 1, which tells that particle 3 can always catch up with particle 1. 
Thus, a classical collision occurs at the singularity point for the triangle loop case.
The condition for this triangle singularity has already been discussed in Ref.~\cite{Bayar:2016ftu}, and here we provide an alternative mathematical method as shown in Eq.(\ref{eq:imaginary}), which will later be used to study the conditions for box singularity.

\section{Two diagrams of the 3-body decay with box single loop}
\label{sec:diagram}

\begin{figure}[htbp]
\centering
\subfigure[]{\includegraphics[width=0.87\linewidth]{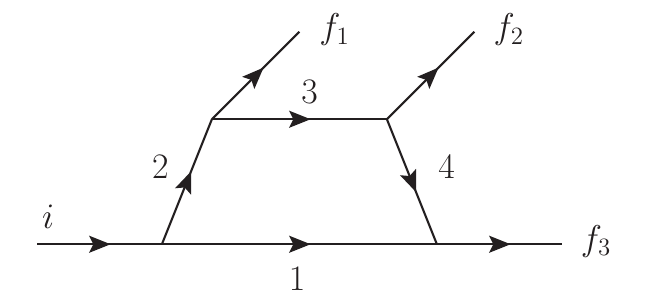} \label{fg:3loopa}}
\subfigure[]{\includegraphics[width=0.87\linewidth]{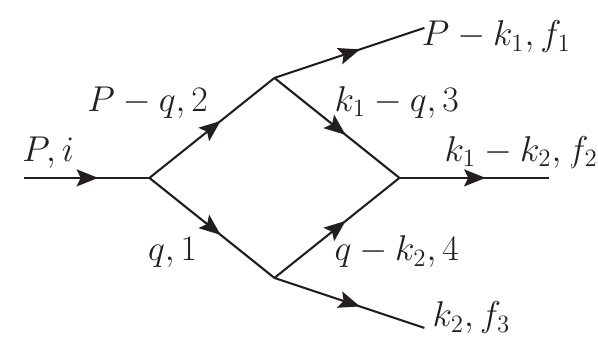} \label{fg:3loopb}}
\caption{Two types of box diagrams. The notation of particles and momentum is shown in (b). \label{fg:3loop}}
\end{figure}

In the previous section, we investigated triangle singularity by analyzing the algebraic properties of the loop integral and obtained a corresponding physical interpretation.
Now we turn to our goal of this paper to explore the singularities in the box loop diagram.
Unlike triangle diagrams that are symmetric, box diagrams have different structures.
There are two different types of box diagrams as shown in Fig.~\ref{fg:3loop}.
Although the loop integrals in these two cases are mathematically identical, the conditions that generate the singularities are different.

In the subsequent calculations, we take the momentum notation $k_2\equiv p_{f_3}$ and $k_1\equiv  p_{f_2}+p_{f_3}$ as given in Fig.~\ref{fg:3loopb}.
Then the loop integral ${\cal I}_3$ of the box diagram writes: 
\begin{align}
  {\cal I}_3 \equiv& \int \frac{{\rm d}^4 q}{ (q^2-m_1^2+i\varepsilon) [(P-q)^2-m_2^2+i\varepsilon] } \nonumber \\
    &\times \frac1{\left[(k_1-q)^2-m_{3}^2+i\varepsilon\right] \left[(k_2-q)^2-m_{4}^2+i\varepsilon\right] }.    
    \label{eq:3loop}
\end{align}
There are a total of eight poles for the integral of $q^0$:
\begin{align}
  q^0 =& \pm \left(\omega_1-i\varepsilon\right), \ M \pm \left(\omega_2-i\varepsilon\right), \nonumber \\ 
  & k_1^0 \pm \left(\omega_3-i\varepsilon\right), \ k_2^0 \pm \left(\omega_4-i\varepsilon\right).
\end{align}
%
%

\begin{figure}[htbp]
\centering
\subfigure[]{\includegraphics[width=0.87\linewidth]{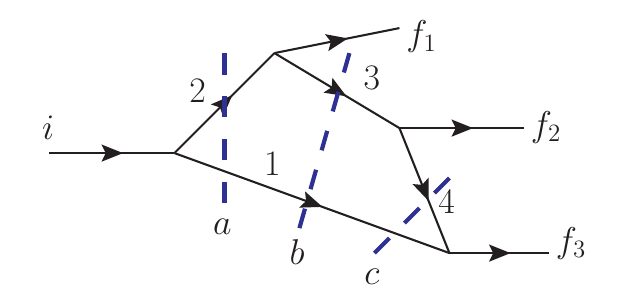} \label{fg:3loopcuta}}
\subfigure[]{\includegraphics[width=0.87\linewidth]{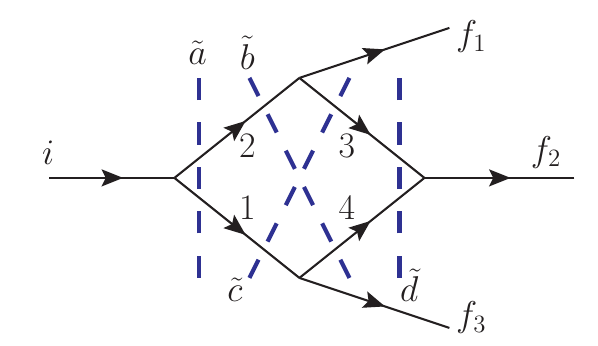} \label{fg:3loopcutb}}
\caption{The relevant cuts for the two types of box diagrams. \label{fg:3loopcut}}
\end{figure}

For the case of Fig.~\ref{fg:3loopa}, we take the poles below the real axis to apply the residue theorem.
It can be found that there are a total of three terms in the denominator that could be zero, and they all come from the residue of pole $\omega_1-i\varepsilon$, which means that only one term ${\cal I}_a$ in ${\cal I}_3$ plays a role when discussing the singularity.
Here, for convenience, we define the three terms that can be zero as:
\begin{align}
a =& M - \omega_1 - \omega_2 + i\varepsilon, \nonumber\\
b =& k_1^0 - \omega_1 - \omega_3 + i\varepsilon, \nonumber\\
c =& k_2^0 - \omega_1 - \omega_4 + i\varepsilon. \label{eq:F1abc}
\end{align}
When these three terms are equal to zero, they correspond to the three cuts as shown in Fig.~\ref{fg:3loopcuta}.
Then we only need to focus on ${\cal I}_a$ to extract the location of the singularity, and it can be written as:
\begin{equation}
{\cal I}_a = \int \frac{{\rm d}^3 \vec{q}}{abc}. 
\label{eq:intera}
\end{equation}

For Fig.~\ref{fg:3loopb}, the situation is more complicated.
There are four terms in the denominator that can be zero, and they are not independent of each other.
These four terms are:
\begin{align}
\tilde{a} =& M-\omega_1 -\omega_2 +i\varepsilon, \nonumber\\
\tilde{b} =& M-k_2^0-\omega_2 -\omega_4 +i\varepsilon, \nonumber\\
\tilde{c} =& k_1^0-\omega_1 -\omega_3 +i\varepsilon, \nonumber\\
\tilde{d} =& k_1^0-k_2^0-\omega_3 -\omega_4 +i\varepsilon.
\label{eq:abcd}
\end{align}
Whether we choose the poles above or below the real axis to apply the residue theorem, there are always two residues that contain such terms. 
It is easy to see that $\tilde{a}+\tilde{d}=\tilde{b}+\tilde{c}$, which means that when any three of them are zero, the remaining one is also zero.
These four terms correspond to the four cuts as given in Fig.~\ref{fg:3loopcutb}.
We choose the residues of poles $M - \omega_2 + i \varepsilon $ and $k_1^0-\omega_3+i\varepsilon$ for subsequent calculations. 
Then ${\cal I}_b$, which actually makes a difference to the singularity, is written as:
\begin{equation}
  {\cal I}_b = \int \mathrm{d}^3{\vec q} \frac{\tilde{a}+\tilde{d}}{\tilde{a}\tilde{b}\tilde{c}\tilde{d}}= \int \mathrm{d}^3{\vec q} \frac{\tilde{b}+\tilde{c}}{\tilde{a}\tilde{b}\tilde{c}\tilde{d}}. \label{eq:interb}
\end{equation}
The detailed calculation of Eq.(\ref{eq:interb}) can be found in Appendix~\ref{app:bform}.

In summary, we obtain two explicit formulas for exploring singularities for the two types of box diagrams.

\section{Singularity conditions in the case of Fig.~\ref{fg:3loopa}}
\label{sec:diagramA}

In this section, we start from Eq.(\ref{eq:intera}) in Fig.~\ref{fg:3loopa} to discuss the conditions for singularity and related properties.

We assume that in the rest frame of the initial particle, $\vec k_2$ is along the $\hat z$-axis, and $\vec k_1$ is in the $\hat x-\hat z$ plane, that is, $\vec k_1=k_1(\sin\theta_1, 0, \cos\theta_1)$ and $\vec k_2=k_2(0, 0,1)$, where $\cos\theta_1=\left(m_{f_2}^2-m_{f_2f_3}^2-m_{f_3}^2+2k_1^0k_2^0\right)/(2k_1k_2)$ with $k_{1,\,2}\ge 0$ and $m_{f_2f_3}$ being the invariant mass of particles $f_2$ and $f_3$. 
The loop integral momentum is defined as $\vec{q}=q(\sin\theta\cos\phi, \sin\theta\sin\phi, \cos\theta)$, and all energies are functions of the integral variables $q$, $\cos\theta$ and $\phi$, written as:
\begin{align}
  \omega_{1,2}(q) =& \sqrt{q^2+m_{1,2}^2}, \nonumber \\
  \omega_3(q,\theta,\phi) =& \big( q^2+k^2_1+m_3^2-2qk_1\cos\theta_1\cos\theta \nonumber \\
  & -2qk_1\sin\theta_1\sin\theta\cos\phi \big)^{\frac12}, \nonumber \\
  \omega_4(q,\theta) =& \sqrt{q^2+k^2_2+m_4^2-2qk_2\cos\theta}.
\end{align}
Then Eq.(\ref{eq:intera}) becomes,
\begin{align}
  {\cal I}_a
  =& \int^\infty_{0}\frac{q^2\mathrm{d}{q}}{M-\omega_1(q)-\omega_2(q)+i\varepsilon} \nonumber\\
  &\times \int^\pi_0 \frac{\mathrm{d}{\cos\theta}}
  {k^0_2-\omega_1(q)-\omega_4(q,\theta)+i\varepsilon} \nonumber\\
  &\times \int^{2\pi}_{0} \frac{\mathrm{d}{\phi}}
  {k^0_1-\omega_1(q)
  -\omega_3(q,\theta,\phi)+i\varepsilon}. \label{eq:interba}
\end{align}

It should be noted that when $\theta_1=0$ or $\pi$, the variable $\phi$ does not appear in the integrand, but Eq.(\ref{eq:interba}) still holds.
We will discuss this case in the second subsection. 
Here we first consider the case where $\theta_1\neq 0$ and $\pi$, and the integral related to $\phi$ can be written as:
\begin{align}
{\cal I}_\phi =& \int^{2\pi}_0 \frac{\mathrm{d}\phi}{k_1^0-\omega_1(q)-\omega_3(q,\theta,\phi)+i\varepsilon} \nonumber \\
=& 2\int^{\pi}_0 \frac{\mathrm{d}\cos\phi \left(k_1^0-\omega_1(q)+\omega_3(q,\theta,\phi)\right)}{\sqrt{1-\cos^2\phi}\left((k_1^0-\omega_1(q))^2-
\omega^2_3(q,\theta,\phi)+i\varepsilon\right)}. \label{eq:iphi000}
\end{align}
We found that only when divergence occurs at the boundary of $\cos\phi$, it diverges after integrating $\phi$.
If the integral concerning $\phi$ does not diverge, then the divergent singularity is, at most, a TS. 
In other words, it tells that $\phi$ should take the boundary value $0$ or $\pi$ when the box singularity happens.
This corresponds to the physical situation where all the internal and external particles involved are in the same $\hat x-\hat z$ plane.
In addition, we need to complete the integration of ${\cal I}_\phi$ around the singular points as:
\begin{align}
{\cal I}^0_\phi =& \int^{\epsilon}_0 \frac{\mathrm{d}\cos\phi \left(k_1^0-\omega_1(q)+\omega_3(q,\theta,\phi)\right)}{\sqrt{1-\cos^2\phi}\left((k_1^0-\omega_1(q))^2- \omega^2_3(q,\theta,\phi)+i\varepsilon\right)} \nonumber \\
\propto & \frac{1}{\sqrt{(k_1^0-\omega_1(q))^2- \omega^2_3(q,\theta,0)+i\varepsilon}}, \nonumber \\
{\cal I}^\pi_\phi =& \int^{\pi}_{\pi-\epsilon} \frac{\mathrm{d}\cos\phi \left(k_1^0-\omega_1(q)+\omega_3(q,\theta,\phi)\right)}{\sqrt{1-\cos^2\phi}\left((k_1^0-\omega_1(q))^2- \omega^2_3(q,\theta,\phi)+i\varepsilon\right)} \nonumber \\
\propto &  \frac{1}{\sqrt{(k_1^0-\omega_1(q))^2- \omega^2_3(q,\theta,\pi)+i\varepsilon}}.
\label{eq:iphipi}
\end{align}

Thus, to extract the singularity condition of Eq.(\ref{eq:intera}), we only need to consider the following integral,
\begin{align}
{\cal I}_a \propto & \int \frac{\mathrm{d}q \ {\cal I}^\prime_{a}}{M-\omega_1(q)-\omega_2(q)+i\varepsilon}, \nonumber \\
{\cal I}^\prime_{a} =& \int \frac{\mathrm{d}\cos\theta }
{(\cos\theta-X(q)+i\varepsilon)\sqrt{\cos(\theta_1\mp\theta)-Y(q)+i\varepsilon}},
\label{eq:iprime1}
\end{align}
where
\begin{align}
X(q) \equiv& \frac{m_4^2-m_1^2-m_{f_3}^2+2k_2^0\sqrt{q^2+m_1^2}}{2k_2q}, \nonumber \\
Y(q) \equiv& \frac{m_3^2-m_{f_2f_3}^2-m_1^2+2k_1^0\sqrt{q^2+m_1^2}}{2k_1q},
\label{eq:xyq}
\end{align}
and the ``$\mp$" sign in the formulas and explanations hereinafter corresponds to the cases where $\phi=0$ and $\pi$, respectively.

The pole of $1/(M-\omega_1(q)-\omega_2(q)+i\varepsilon)$ is $q_{\rm on}+i \varepsilon$.
Similar to the case of triangle singularity, we know that $q_{\rm on}-i \varepsilon$ also needs to be a pole of ${\cal I}^\prime_{a}$, i.e. ${\cal I}^\prime_{a}(q_{\rm on}-i \varepsilon)=\infty$.
Then we turn to the integral of $\cos\theta$ in ${\cal I}^\prime_{a}$.
The first denominator of ${\cal I}^\prime_{a}$ contains only one first-order pole, that is, $\cos\theta=X(q)-i\varepsilon$, and further derivation is required for the second denominator.

In order to give a explicit solution for $\cos(\theta_1\mp\theta)-Y(q)+i\varepsilon = 0$, we define the following two functions:
\begin{align}
Y^+_0(q) \equiv& Y(q)\cos\theta_1 + \sqrt{1-Y^2(q)}\sin\theta_1, \nonumber \\
Y^-_0(q) \equiv& Y(q)\cos\theta_1 - \sqrt{1-Y^2(q)}\sin\theta_1.
\label{eq:Y0Z0pm}
\end{align} 
After the derivation given in Appendix~\ref{app:Y0pm}, we obtain the solutions as:
\begin{align}
\cos\theta = 
\left\{
\arraycolsep=4pt\def\arraystretch{1.1}
\begin{array}{ll}
Y^+_0(q)+i\varepsilon_0, & \phi=0\, \&\, \theta_1>\theta>0 \\
Y^-_0(q)-i\varepsilon_0, & \phi=0\, \&\, \theta_1<\theta<\pi  \\
Y^+_0(q)-i\varepsilon_\pi,& \phi=\pi\, \&\, \pi-\theta_1>\theta>0 \\
Y^-_0(q)+i\varepsilon_\pi,& \phi=\pi\, \&\, \pi-\theta_1<\theta<\pi 
\end{array}
\right.,
\label{eq:costheta}
\end{align}
where the imaginary parts are:
\begin{align}
\varepsilon_0 =& \frac{\varepsilon\sin\theta}{|\sin(\theta_1-\theta)|}>0, \nonumber \\
\varepsilon_\pi =& \frac{\varepsilon\sin\theta}{|\sin(\theta_1+\theta)|}>0. 
\end{align} 
Furthermore, for the case where $\cos\theta$ is at the boundary, there are special solutions that correspond to triangle singularities rather than box singularities. 
Their specific forms and related discussions are also given in Appendix~\ref{app:Y0pm}.

Then there are only two situations that make ${\cal I}^\prime_a$ diverge at $q=q_{\rm on}-i\varepsilon$.
Firstly, the momentum directions of the three intermediate particles $1$, $3$, and $4$ are different in the rest frame of the mother particle.
Moreover, in the rest frame of the final particles $f_2$ and $f_3$, the directions of particles 1 and 4 are also different.
In this case, the singularity happens when the integral path of $\cos\theta$ is pinched between the two poles from two denominators of ${\cal I}^\prime_a$.
Since the singularity generated here has nothing to do with triangle singularity, we refer to such singularity as the general box singularity.
Secondly, all three final particles are moving on the same line, i.e. $\theta_1=0$ or $\pi$.
In this case, the entire integration is independent of the variable $\phi$. 
We will later find that even if the two $\cos\theta$ poles of ${\cal I}^\prime_a$ can be pinched when $\theta_1=\pi$, ${\cal I}^\prime_a$ is still not divergent at $q_{\rm on}-i\varepsilon$. 
Thus, the possible singularity only arises from boundary divergence, which results in all particles moving on the same line.
We will discuss these two cases in detail in the subsequent subsections.

\subsection{The general box singularity}
\label{sec:noTSA}

In this subsection, we consider the case that the integral path of $\cos\theta$ is pinched between the two poles from the two denominators of ${\cal I}^\prime_a$ in Eq.(\ref{eq:iprime1}). 
Since we already have the pole $X(q)-i\varepsilon$ of the first denominator, which lies below the real $\cos\theta$-axis, another $\cos\theta$ pole from the second denominator should be above the real axis, which are the solutions with positive imaginary parts in Eq.(\ref{eq:costheta}), i.e.,
\begin{align}
\cos\theta = 
\left\{
\arraycolsep=3pt\def\arraystretch{1.1}
\begin{array}{ll}
Y^+_0(q)+i\varepsilon_0, &  \phi=0\, \&\, \theta_1>\theta>0 \\
Y^-_0(q)+i\varepsilon_\pi,&  \phi=\pi\, \&\, \pi-\theta_1<\theta<\pi 
\end{array}
\right. .
\label{eq:costheta1}
\end{align}
Therefore, for the cases of $\phi=0$ and $\pi$, there are restrictions $\theta_1 > \theta$ and $\theta_1 +\theta > \pi$ respectively.
Notably, these two restrictions $\theta_1 > \theta$ for $\phi=0$ and $\theta_1+\theta > \pi$ for $\phi=\pi$ can be reduced to the same constraint $\vec v_{3\perp1} \cdot \vec v_{4\perp1} < 0$, where $\vec v_{3\perp1}$ and $\vec v_{4\perp1}$ are the velocities of intermediate particles 3 and 4 perpendicular to the direction of particle $1$.
For convenience, we take $ v_{3\perp1}>0>v_{4\perp1}$ in the following calculation.
Besides, we can understand this condition from a physical perspective that particles $3$ and $4$ move in opposite directions perpendicular to particle 1. 

For simplicity, we define:
\begin{align}
Y_0(q) = \cos\theta_{\text{on}} \equiv
\left\{
\arraycolsep=3pt\def\arraystretch{1.1}
\begin{array}{ll}
Y^+_0(q),&  \phi=0  \\
Y^-_0(q),&  \phi=\pi  
\end{array}
\right. .
\label{eq:Y0}
\end{align}
Then the two poles of the integrand are $X(q)-i\varepsilon$ and $Y_0(q)+i\varepsilon$, and we can write ${\cal I}^\prime_a$ in the form as:
\begin{align}
  {\cal I}^\prime_a \propto & \int \frac{\mathrm{d}\cos\theta }{(\cos\theta - X(q) +i\varepsilon)\sqrt{\cos\theta - Y_0(q) -i\varepsilon}} \nonumber\\
  \propto & \frac{1}{\sqrt{X(q)-Y_0(q)-i\varepsilon}}.
  \label{eq:Iaprime}
\end{align}
It can be seen that ${\cal I}^\prime_a$ is divergent only when 
\begin{equation}
    X(q_{\rm on}-i \varepsilon)-Y_0(q_{\rm on}-i \varepsilon)-i\varepsilon^\prime=0. 
    \label{Eq:qonam}
\end{equation}
We find that the real part of Eq.(\ref{Eq:qonam}) naturally holds when all the intermediate particles are on-shell, i.e. ${\rm Re}[X(q_{\rm on})]={\rm Re}[Y_0(q_{\rm on})]$. 
So, we only need to focus on the imaginary part of Eq.(\ref{Eq:qonam}).
It is clear that imaginary part of $X(q_{\rm on}-i \varepsilon)-Y_0(q_{\rm on}-i \varepsilon)$ should be positive to hold Eq.(\ref{Eq:qonam}). 
By first-order expansion of $\varepsilon$, we get:
\begin{equation}
  \frac{{\rm d}X}{{\rm d}q} \Big|_{q\to q_{\rm on}} - \frac{{\rm d}Y_0}{{\rm d}Y}\frac{{\rm d}Y}{{\rm d}q}\Big|_{q\to q_{\rm on}} < 0.
  \label{eq:imqonam}
\end{equation}
It results in (the detailed derivation can be found in Appendix~\ref{app:condition}):
\begin{equation}
  - \frac{v_1-v_{4\parallel1}}{v_{4\perp1}} + \frac{v_1-v_{3\parallel1}}{v_{3\perp1}}< 0. 
  \label{eq:condition}
\end{equation}
where
\begin{align}
v_1 =& \frac{q_{\rm on}}{\omega_1(q_{\rm on})}, \nonumber \\
v_{3\perp1} =&|\vec{v}_{3\perp1}| =
\left\{
\arraycolsep=3pt\def\arraystretch{1.3}
\begin{array}{ll}
\frac{k_1\sin(\theta_1-\theta_{\rm on})}{\omega_3(q_{\rm on},\theta_{\rm on},0)},&  \phi=0 \nonumber \\
\frac{k_1\sin(\theta_1+\theta_{\rm on}-\pi)}{\omega_3(q_{\rm on},\theta_{\rm on},\pi)},&  \phi=\pi
\end{array}
\right. , \nonumber \\
v_{4\perp1} =& -|\vec{v}_{4\perp1}| =-\frac{k_2\sin\theta_{\rm on}}{\omega_4(q_{\rm on},\theta_{\rm on})}.
\label{eq:v134}
\end{align}
The signs of $v_{3\parallel1}$ and $v_{4\parallel1}$ are undetermined and they are defined as:
\begin{align}
v_{3\parallel1} =&
\left\{
\arraycolsep=3pt\def\arraystretch{1.3}
\begin{array}{ll}
\frac{k_1\cos(\theta_1-\theta_{\rm on})-q_{\rm on}}{\omega_3(q_{\rm on},\theta_{\rm on},0)},&  \phi=0 \\
\frac{k_1\cos(\theta_1+\theta_{\rm on})-q_{\rm on}}{\omega_3(q_{\rm on},\theta_{\rm on},\pi)},&  \phi=\pi  
\end{array}
\right. , \nonumber \\
v_{4\parallel1} =& \frac{k_2\cos \theta_{\rm on}-q_{\rm on}}{\omega_4(q_{\rm on},\theta_{\rm on})}.
\label{eq:v4p1}
\end{align}

\begin{figure}[htbp]
\centering
  \includegraphics[width=0.8\linewidth]{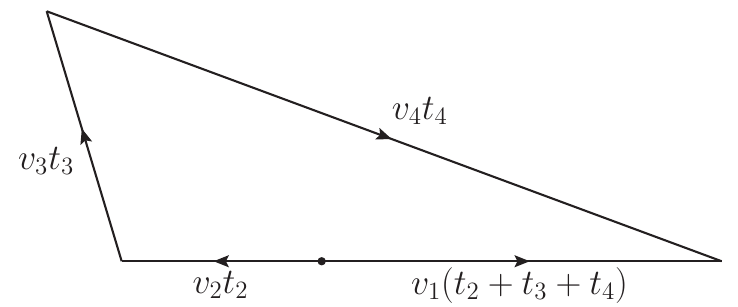}
  \caption{The diagram of the internal particle trajectories in real space-time when box singularity happens in the case of Fig.~\ref{fg:3loopa}.}
  \label{Fig:pattern1}
\end{figure}

Eq.(\ref{eq:condition}) can be explained physically.
Assume that the lifetimes of intermediate particles $2$, $3$ and $4$ are $t_2$, $t_3$ and $t_4$, respectively.
For particle 4 to catch up with particle 1 and collide as shown in Fig.~\ref{Fig:pattern1}, two conditions must be met:
\begin{align}
v_{3\perp1}t_3 =&-v_{4\perp1}t_4, \nonumber \\
v_2t_2+v_{3\parallel1}t_3+v_{4\parallel1}t_4 =& v_1(t_2+t_3+t_4).
\end{align}
Since $v_1$ is positive and $v_2$ is negative, we have:
\begin{align}
 (v_1-v_2)t_2 =& v_{3\parallel1}t_3+v_{4\parallel1}t_4-v_1(t_3+t_4) \nonumber \\
  =& \left(\frac{v_{3\parallel1}-v_1}{v_{3\perp1}} - \frac{v_{4\parallel1}-v_1}{v_{4\perp1}}\right)v_{3\perp1}t_3 > 0. 
  \label{Eq:pattern1}
\end{align}
It is exactly the same as Eq.(\ref{eq:condition}).
Therefore, the physical explanation of Eq.(\ref{eq:condition}) is that particle 4 can catch up with particle 1 to collide, which is consistent with the statement of Coleman-Norton theorem in Ref.~\cite{Coleman:1965xm}.

In summary, for the case where the integral path of $\cos\theta$ is pinched between two poles, we obtain the following constraints:
\begin{align}
\vec v_{3\perp1} &\cdot \vec v_{4\perp1} <0, \nonumber \\
\frac{v_{4\parallel1}-v_1}{v_{4\perp1}} &- \frac{v_{3\parallel1}-v_1}{v_{3\perp1}} < 0, 
\label{eq:conditionlast}
\end{align}
and all intermediate particles are on-shell.
The corresponding physical situation is that particle 4 can catch up with particle 1 to collide in real space-time.

\subsection{For the case of \texorpdfstring{$\theta_1=0$}{theta1=0} or \texorpdfstring{$\pi$}{pi}}
\label{sec:olineA}

In this subsection, we study the situation where all final particles move in the same straight line, i.e. $\theta_1=0$ or $\pi$.
Since the integral variable $\phi$ is independent of the integrand in Eq.(\ref{eq:intera}), we can obtain:
\begin{align}
{\cal I}_a =& \int \frac{\mathrm{d}q \ {\cal \tilde{I}}^\prime_{a}}{M-\omega_1(q)-\omega_2(q)+i\varepsilon}, \label{eq:iaa} \\
{\cal \tilde{I}}^\prime_{a} =&  \int \frac{\mathrm{d}\cos\theta}{(k_2^0-\omega_1(q)-\omega_4(q,\theta)+i\varepsilon)} \nonumber \\
& \times \frac1{(k_1^0-\omega_1(q)-\omega_3(q,\theta)+i\varepsilon)} \nonumber \\
=& \int \frac{\mathrm{d}\cos\theta \tilde{f}(q,\theta)}{(\cos\theta-X(q)+i\varepsilon)(\pm\cos\theta-Y(q)+i\varepsilon)},
\label{eq:iabline}
\end{align}
where $\omega_3(q,\theta)=\sqrt{q^2+m_3^2+k_1^2\pm2qk_1\cos\theta}$.
The ``$\pm$'' sign in Eq.(\ref{eq:iabline}) corresponds to $\theta_1=0$ and $\pi$ respectively, $X(q)$ and $Y(q)$ are defined in Eq.(\ref{eq:xyq}), and
\begin{align}
\tilde{f}(q,\theta) = \frac{(k_2^0-\omega_1(q)+\omega_4(q,\theta))(k_1^0-\omega_1(q)+\omega_3(q,\theta))}{4q^2k_1k_2}
\end{align}
is just a regular function without any singularity, which we will drop when studying the divergence conditions.

Since $q=q_{\rm on}+i\varepsilon$ is a zero point in the denominator of Eq.(\ref{eq:iaa}), in order to make ${\cal I}_a$ diverge, we need to ensure that ${\cal \tilde{I}}^\prime_a$ diverges at $q=q_{\rm on}-i\varepsilon$ after integrating $\cos\theta$.
Similarly, there are two possible scenarios: the integral path is pinched by two $\cos\theta$ poles, or there exists an endpoint divergence.

We first prove that ${\cal \tilde{I}}^\prime_{a}$ cannot diverge at $q_{\rm on}-i\varepsilon$ due to the integral path being pinched between the two $\cos\theta$ poles.
This is obvious for $\theta_1=0$, as the poles from both denominators of Eq.(\ref{eq:iabline}) are located below the real axis of $\cos\theta$.
For $\theta_1=\pi$, we can obtain a pole $\cos\theta = -Y(q_{\rm on})+i\varepsilon$ that is always above the real axis.
In this case, even if the two $\cos\theta$ poles can be pinched, $q=q_{\rm on}-i\varepsilon$ cannot be the pole of ${\cal I}^\prime_{a}$. 
The derivation is also provided in Appendix~\ref{app:condition}.

Therefore, only the case of endpoint divergence can make ${\cal \tilde{I}}^\prime_{a}$ diverge at $q_{\rm on}-i\varepsilon$, i.e. $\theta=0$ or $\pi$.
At this point, all particles are moving on the same line.
Obviously, there are four possibilities for $\theta_1$ and $\theta$ to be $0$ or $\pi$.
We will discuss these four cases one by one.

\subsubsection{\texorpdfstring{$\theta_1=\theta=\pi$}{theta1=theta=pi}}

For $\theta_1=\theta=\pi$, ${\cal \tilde{I}}^\prime_{a}$ can be calculated as:
\begin{align}
{\cal \tilde{I}}^\prime_{a} \propto& \int \frac{\mathrm{d}\cos\theta }{(\cos\theta-(X(q)-i\varepsilon_X))(\cos\theta+(Y(q)-i\varepsilon_Y))} \nonumber \\
\propto& \frac{\log(-1-(X(q)-i\varepsilon_X))-\log(-1+(Y(q)-i\varepsilon_Y))}{Y(q)+X(q)-i(\varepsilon_X+\varepsilon_Y)}.
\label{eq:Iap}
\end{align}
We first check the divergence condition when $q=q_{\text{on}}-i\varepsilon$, and there are:
\begin{align}
X(q_{\text{on}}-i\varepsilon)=&X(q_{\text{on}})-i\varepsilon\left.\frac{\mathrm{d} X(q)}{q}\right|_{q\to q_{\text{on}}} \nonumber \\
=&X(q_{\text{on}})-\frac{i\varepsilon}{q_{\text{on}}}\left(\frac{q_{\text{on}}/\omega_1(q_{\rm on})}{k_2/k^0_2}-X(q_{\text{on}})\right), \nonumber \\
Y(q_{\text{on}}-i\varepsilon)=&Y(q_{\text{on}})-i\varepsilon\left.\frac{\mathrm{d} Y(q)}{q}\right|_{q\to q_{\text{on}}} \nonumber \\
=&Y(q_{\text{on}})-\frac{i\varepsilon}{q_{\text{on}}}\left(\frac{q_{\text{on}}/\omega_1(q_{\rm on})}{k_1/k^0_1}-Y(q_{\text{on}})\right).
\label{eq:iablineXY}
\end{align}
In this case, we have $X(q_{\text{on}})=-1$ and $Y(q_{\text{on}})=1$ for the condition of all intermediate particle on the mass shell, then the first $\log$ term in Eq.(\ref{eq:Iap}) is:
\begin{align}
&(-1-(X(q_{\rm on}-i\varepsilon)-i\varepsilon_X)) \nonumber \\
=& \frac{i\varepsilon}{q_{\text{on}}}
 \left(\frac{q_{\text{on}}/\omega_1(q_{\rm on})}{k_2/k^0_2}+1\right) +i\varepsilon_X \nonumber \\
>& 0.
\end{align}
It does not diverge at $q=q_{\text{on}}-i\varepsilon$.
For the second $\log$ term, we have:
\begin{align}
&(-1+(Y(q_{\rm on}-i\varepsilon)-i\varepsilon_Y)) \nonumber \\
=& - \frac{i\varepsilon}{q_{\text{on}}}
 \left(\frac{q_{\text{on}}/\omega_1(q_{\rm on})}{k_1/k^0_1}-1\right) -i\varepsilon_Y \nonumber \\
=&-\frac{i\varepsilon}{q_{\text{on}}}\frac{v_3-v_1}{k_1/\omega_3(q_{\rm on},\pi)}-i\varepsilon_Y,
\label{eq:logYpipi}
\end{align}
%
where $v_1=-q_{\text{on}}/\omega_1<0$ and $v_3=(q_{\rm on}-k_1)/\omega_3(q_{\rm on}, \pi)$.
When $v_3<v_1<0$, Eq.(\ref{eq:logYpipi}) can be zero, telling that $q_{\text{on}}-i\varepsilon$ can be a pole of ${\cal \tilde{I}}^\prime_{a}$.
For the denominator of Eq.(\ref{eq:Iap}), we have:
\begin{align}
&Y(q)+X(q)-i(\varepsilon_X+\varepsilon_Y) \nonumber \\
=& -\frac{i\varepsilon}{q_{\text{on}}}
\left(\frac{q_{\text{on}}/\omega_1(q_{\rm on})}{k_2/k^0_2}
+\frac{q_{\text{on}}/\omega_1(q_{\rm on})}{k_1/k^0_1}\right)
-i(\varepsilon_Y+\varepsilon_X) \nonumber \\
<& 0,
\end{align}
which will not have a pole of $q_{\text{on}}-i\varepsilon$.
   
In summary, we find that divergence occurs only in the $\log\left(-1+(Y(q)-i\varepsilon_Y)\right)$ term with $v_3<v_1<0$ such that $\left(-i\varepsilon\frac{\omega_3(q_{\rm on},\pi)}{q_{\text{on}}k_1}(v_3-v_1)-i\varepsilon_Y)\right)\to 0$.
Then ${\cal I}_{a}$ can be expressed as:
\begin{align}
{\cal I}_{a} \propto&\int \mathrm{d}q \frac{\log(q-q_{\text{on}}-i\varepsilon_1)-\log(q-q_{\text{on}}+i\varepsilon)}{(q-q_{\text{on}}-i\varepsilon_2)(q-q_{\text{on}}-i\varepsilon_3)}.
\label{eq:iablinepipia}
\end{align}

\begin{figure}[htbp]
 \centering
 \includegraphics[width=0.93\linewidth]{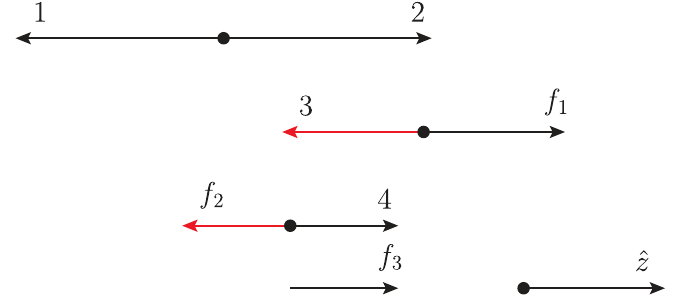}
 \caption{The directions of each particle in the case of $\theta_1=\theta=\pi$. Black line: the direction is determined; red line: the direction is determined only when singularity occurs.}
 \label{Fig:pipi}
\end{figure}

We then turn to the physics of this situation.
When the singularity occurs, we find that the particles $2$, $4$, $f_1$, and $f_3$ are moving along the $\hat z$-axis, while particles $1$, $3$ and $f_2$ are moving in the opposite direction, as shown in Fig.~\ref{Fig:pipi}.
The singularity condition also restricts $|v_1| < |v_3|$, which indicates that particle 3 can collide with particle 1 and generate particles $f_2$ and $f_3$ by exchanging particle $4$.
Thus, this singularity is simply due to the triangle singularity among intermediate particles $1$, $2$, and $3$, as well as an on-shell particle $4$. 
Note that in this case, particle $4$ actually has no chance to ``collide'' with particle $1$.
It appears to conflict with the statement of Coleman-Norton theorem, which we will discuss at the end of this subsection since it will happen again in the latter cases.

\subsubsection{\texorpdfstring{$\theta_1=\pi$}{theta1=pi} and \texorpdfstring{$\theta=0$}{theta=0}}

\begin{figure}[htbp]
 \centering
 \includegraphics[width=0.93\linewidth]{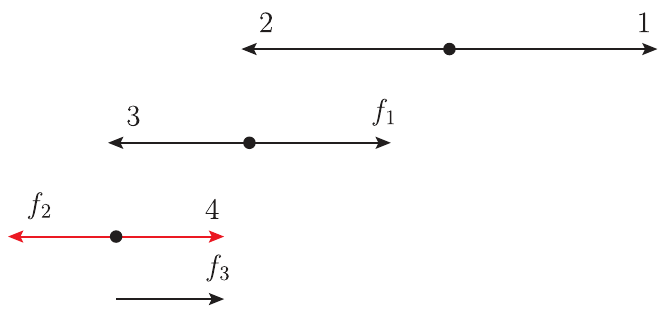}
 \caption{The directions of each particle in the case of $\theta_1=\pi$ and $\theta=0$. Black line: the direction is determined; red line: the direction is determined only when singularity occurs.}
 \label{Fig:pi0}
\end{figure}

For $\theta_1=\pi$ and $\theta=0$, we have $X(q_{\text{on}})=1$, $Y(q_{\text{on}})=-1$, $v_1=q_{\text{on}}/\omega_1(q_{\rm on}) > 0$, $v_3=(-k_1-q_{\rm on})/\omega_3(q_{\rm on}, 0)<0$ and $v_4=(k_2-q_{\rm on})/\omega_4(q_{\rm on}, 0)$ when all intermediate particles are on the mass shell.
%
Then ${\cal \tilde{I}}^\prime_a$ can be derived with $q\to q_{\text{on}}-i\varepsilon$ as:
\begin{align}
{\cal \tilde{I}}^\prime_{a} \propto&\int \frac{\mathrm{d}\cos\theta }{(\cos\theta-(X(q)-i\varepsilon_X))(\cos\theta+(Y(q)-i\varepsilon_Y))} \nonumber \\
\propto&\frac{\log(1-(X(q)-i\varepsilon_X))-\log(1+(Y(q)-i\varepsilon_Y))}{Y(q)+X(q)-i(\varepsilon_X+\varepsilon_Y)}.
\label{eq:iablinepi0}
\end{align}
It can be found that divergence only appears in the $\log\left(1-(X(q_{\rm on}-i\varepsilon)-i\varepsilon_X)\right)$ term when $v_4>v_1>0$, because at this point,
\begin{align}
&1-(X(q_{\rm on}-i\varepsilon)-i\varepsilon_X) \nonumber \\
=& \frac{i\varepsilon}{q_{\text{on}}} \left(\frac{q_{\text{on}}/\omega_1(q_{\rm on})}{k_2/k^0_2}-1\right) +i\varepsilon_X \nonumber \\
=& i\varepsilon\frac{\omega_4(q_{\rm on},0)}{q_{\text{on}}k_2}(v_1-v_4)+i\varepsilon_X,
\label{eq:iablinepi02}
\end{align}
can be zero.
Thus, similar to Eq.(\ref{eq:iablinepipia}), ${\cal I}_{a}$ becomes:
\begin{align}
{\cal I}_{a} \propto \int \mathrm{d}q 
\frac{\log(q-q_{\text{on}}+i\varepsilon_1)-\log(q-q_{\text{on}}-i\varepsilon)}{(q-q_{\text{on}}-i\varepsilon_2)(q-q_{\text{on}}-i\varepsilon_3)}.
\label{eq:iablinepi0a}
\end{align}
We present the momenta with determined directions in Fig.~\ref{Fig:pi0}.
It can be seen that the singularity occurs only when the direction of particle 4 is along the $\hat z$-axis and $v_1 < v_4$, which corresponds to the situation that particle 4 can collide with particle 1.
In this case, a classical collision occurs in reality.

\subsubsection{\texorpdfstring{$\theta_1=\theta=0$}{theta1=theta=0}}

In the case of $\theta_1=\theta=0$, there are $X(q_{\text{on}})=1$, $Y(q_{\text{on}})=1$, $v_1=q_{\text{on}}/\omega_1 > 0$, $v_3=(k_1-q_{\rm on})/\omega_3(q_{\rm on}, 0)$ and $v_4=(k_2-q_{\rm on})/\omega_4(q_{\rm on}, 0)$.
%
Thus, for ${\cal \tilde{I}}^\prime_a$, we have: 
%
\begin{align}
{\cal \tilde{I}}^\prime_{a} \propto&\int \frac{\mathrm{d}\cos\theta }{(\cos\theta-(X(q)-i\varepsilon_X))(\cos\theta-(Y(q)-i\varepsilon_Y))} \nonumber \\
\propto&\frac{\log(1-(X(q)-i\varepsilon_X))-\log(1-(Y(q)-i\varepsilon_Y))}{X(q)-Y(q)-i(\varepsilon_X-\varepsilon_Y)}.
  \label{eq:iabline00a}
\end{align}
This case is different from the previous two cases in that the two $\log$ terms can diverge at $q=q_{\text{on}}-i\varepsilon$ when $v_3>v_1$ and $v_4 > v_1$, respectively.
When $v_3>v_1$, ${\cal I}_{a}$ can be obtained by Eq.(\ref{eq:iablinepipia}), and when $v_4>v_1$, it is as shown in Eq.(\ref{eq:iablinepi0a}).

\begin{figure}[htbp]
 \centering
 \includegraphics[width=0.93\linewidth]{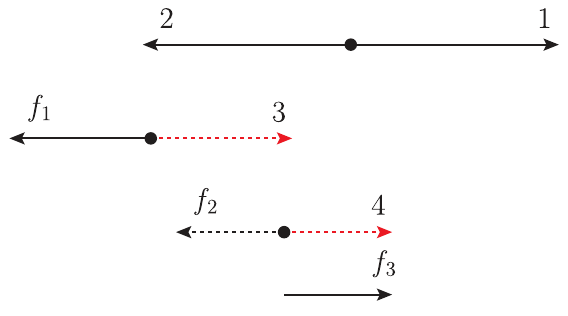}
 \caption{The directions of each particle in the case of $\theta_1=\theta=0$. Black solid line: the direction is determined; black dashed line: the direction is uncertain and is only drawn as an example; red dashed line: when singularity occurs, at least one follows the given direction, and the direction of the other is uncertain.} 
 \label{Fig:00}
\end{figure}

At this point, only four particles $1$, $2$, $f_1$, and $f_3$ have definite directions according to the values of the two angles, as shown in Fig.~\ref{Fig:00}.
The two possible singularity conditions $v_3>v_1$ and $v_4 > v_1$ can correspond to the physical situations that particles $3$ and $4$ can ``catch up'' with particle $1$, respectively.
Therefore, singularity condition $v_3>v_1$ is similar to the case of $\theta_1=\theta=\pi$, and the condition $v_4>v_1$ is similar to the case of $\theta_1=\pi$ and $\theta=0$.
The difference is that there is a possibility that both $v_3>v_1$ and $v_4 > v_1$ can be satisfied simultaneously, while in the previous two cases only one condition can be met.  
Again, if $v_4 > v_1$ is not satisfied, the situation is similar to the first case in that a "collision" between particle 4 and 1 does not happen.
We will discuss this at the end of this subsection.

\subsubsection{\texorpdfstring{$\theta_1=0$}{theta1=0} and \texorpdfstring{$\theta=\pi$}{theta=pi}}
\label{sec:0pia}

\begin{figure}[htbp]
 \centering
 \includegraphics[width=0.93\linewidth]{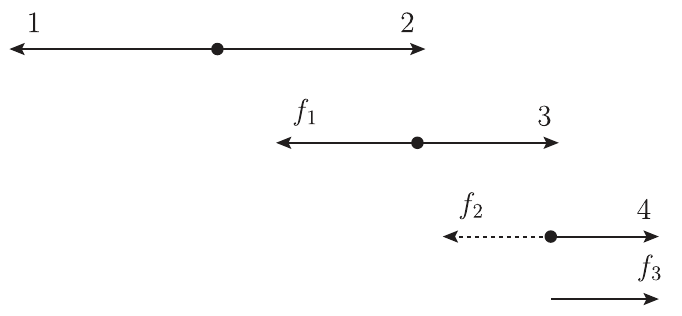}
 \caption{The directions of each particle in the case of $\theta_1=0$ and $\theta=\pi$. Solid line: the direction is determined; dashed line: the direction is uncertain and is only drawn as an example.} 
 \label{Fig:0pi}
\end{figure}

The last case is $\theta_1=0$ and $\theta=\pi$, and we can get $X(q_{\text{on}})=-1$, $Y(q_{\text{on}})=-1$, $v_1=-q_{\text{on}}/\omega_1 < 0$, $v_3=(k_1+q_{\rm on})/\omega_3(q_{\rm on}, 0)>0$ and $v_4=(k_2+q_{\rm on})/\omega_4(q_{\rm on}, 0)>0$.
%
Then we have: 
%
\begin{align}
{\cal \tilde{I}}^\prime_{a} \propto& \int \frac{\mathrm{d}\cos\theta }{(\cos\theta-(X(q)-i\varepsilon_X))(\cos\theta-(Y(q)-i\varepsilon_Y))} \nonumber \\
\propto& \frac{\log(-1-(X(q)-i\varepsilon_X))-\log(-1-(Y(q)-i\varepsilon_Y))}{X(q)-Y(q)-i(\varepsilon_X-\varepsilon_Y)}.
\label{eq:org}
\end{align}
Both $\log$ terms are always positive in this case when $q\to q_{\text{on}}-i\varepsilon$, while the denominator has not been determined due to its correlation with the relative sizes of $\varepsilon_X$ and $\varepsilon_Y$.
Starting from Eq.(\ref{eq:org}), by setting $\delta\equiv X(q)-Y(q)-i(\varepsilon_X-\varepsilon_Y)$, we can obtain:
\begin{align}
{\cal \tilde{I}}^\prime_{a} \propto& \frac{\log(-1-X(q)+i\varepsilon_X)-\log(-1-X(q)+i\varepsilon_X+\delta)}{\delta} \nonumber \\
\propto& \frac{1}{-1-(X(q)-i\varepsilon_X)}.
\end{align}
Here we take $\delta \ll \left|-1-(X(q)-i\varepsilon_X)\right|$, since $\delta$ is the difference between the two terms in the $\log$,  which are both close to zero. 
Obviously, $q_{\text{on}}-i\varepsilon$ is not a pole of ${\cal \tilde{I}}^\prime_{a}$, which also shows that there is no box singularity.
The momentum directions of all relevant particles except $f_2$ are determined, as shown in Fig.~\ref{Fig:0pi}.
We can see that in this case particles 3 and 4 would never collide with particle 1.

To sum up, the physical condition for box singularity with all particles on the same line is that all intermediate particles are on-shell, and particles 3 or 4 can ``catch up'' with particle 1, that is, the velocities of particle 3 or 4 must be larger than that particle 1 and in the same direction.
An interesting situation is that particle 3 can catch up with particle 1 but particle 4 cannot collide with particle 1, that is, the decay of $3\to 4+f_2$ occurs before particle 3 catches up with particle 1, so particle 4 cannot collide with particle 1 in real space-time. 
The classical collision does not occur, which seems to be inconsistent with the description of Coleman-Norton theorem.
In fact, one can not determine the lifetime of the intermediate particles in a Feynman diagram, even if they are on-shell, so the above scenario is just one possibility.
Another plausible scenario is that particle 3 can live long enough to surpass particle 1 before decaying into particle 4 and $f_2$. 
Then, since particle 4 is moving slower than particle 1, particle 1 will ``catch up'' with particle 4, and such a process can still be regarded as a classical collision.
Therefore, regarding the statement of Coleman-Norton theorem, we interpret it as ``classical collisions'' should be possible when Landau pole exists.

\section{Singularity conditions in the case of Fig.~\ref{fg:3loopb}}
\label{sec:diagramB}

\begin{figure}[htbp]
\centering
\subfigure[]{\includegraphics[width=0.87\linewidth]{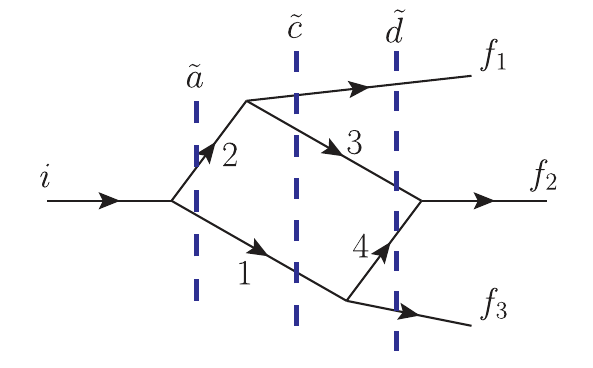} \label{fg:baba}}
\subfigure[]{\includegraphics[width=0.87\linewidth]{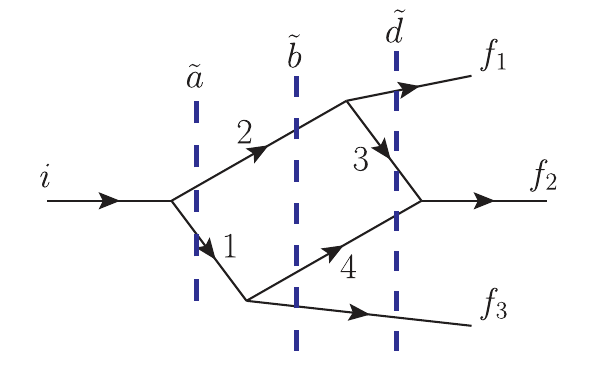} \label{fg:babb}}
\caption{The relevant cuts for two terms of ${\cal I}_b$.
\label{fg:bab}}
\end{figure}

In this section, we consider the process shown in Fig.~\ref{fg:3loopb}.
We start from Eq.(\ref{eq:interb}) to study the singularity, that is,
\begin{align}
{\cal I}_b = {\cal I}_{b_1}+ {\cal I}_{b_2} = \int \mathrm{d}^3{\vec q} \frac1{\tilde{a}\tilde{c}\tilde{d}} + \int \mathrm{d}^3{\vec q} \frac{1}{\tilde{a}\tilde{b}\tilde{d}}.
\end{align}
The denominators of ${\cal I}_{b_1}$ and ${\cal I}_{b_2}$ correspond to the cuts in Fig.~\ref{fg:baba} and Fig.~\ref{fg:babb}, respectively.
They are actually symmetrical to each other and can be obtained by simultaneously exchanging particles 1 and 2, 3 and 4, and $f_1$ and $f_3$.
Thus, once we have the conditions for the singularity of ${\cal I}_{b_1}$, the conditions for the singularity of ${\cal I}_{b_2}$ can naturally be obtained by exchanging the above particle indices.

Mathematically, we make a different momentum assumption than in the case of Fig.~\ref{fg:3loopa}, i.e. $\vec{k}_1=k_1(0,0,1)$, $\vec{k}_2=k_2(\sin \theta_1, 0, \cos\theta_1)$ and $\vec{q}=q(\sin\theta\cos\phi, \sin\theta\sin\phi, \cos\theta)$.
The value of $\cos\theta_1$ is still $\left(m_{f_2}^2-m_{f_2f_3}^2-m_{f_3}^2+2k_1^0k_2^0\right)/(2k_1k_2)$.
The energies $\omega_i$ are:
\begin{align}
  \omega_{1,2}(q)=& \sqrt{q^2+m_{1,2}^2}, \nonumber \\
  \omega_3(q,\theta)=& \sqrt{q^2+k^2_1+m_3^2-2qk_1\cos\theta}, \nonumber \\
  \omega_4(q,\theta,\phi)=& (q^2+k^2_2+m_4^2-2qk_2\cos\theta_1\cos\theta \nonumber \\
 & -2qk_2\sin\theta_1\sin\theta\cos\phi)^{\frac12}.
\label{eq:omega}
\end{align}
Then we can write ${\cal I}_{b_1}$ as:
\begin{align}
{\cal I}_{b_1} =& \int\frac{q^2\mathrm{d}{q}}{M-\omega_1(q)-\omega_2(q)+i\varepsilon} \nonumber \\
  &\times \int\frac{\mathrm{d}{\cos\theta}}{k^0_1-\omega_1(q)-\omega_3(q,\theta)+i\varepsilon} \nonumber \\
  &\times \int\frac{\mathrm{d}{\phi}}{k^0_1-k^0_2-\omega_3(q,\theta) - \omega_4(q,\theta,\phi)+i\varepsilon}.
\label{eq:interIb1}
\end{align}
Likewise, $\phi$ is independent of the integrand when $\theta_1=0$ or $\pi$.
These cases will be discussed separately in the second subsection, and here we only consider the case of $\theta_1\neq 0$ and $\pi$.

Similar to Eqs.(\ref{eq:iphi000}) and (\ref{eq:iphipi}), we can obtain:
\begin{align}
{\cal I}^b_\phi =& \int\frac{\mathrm{d}{\phi}}{k^0_1-k^0_2-\omega_3(q,\theta) - \omega_4(q,\theta,\phi)+i\varepsilon} \nonumber \\
\propto& \frac1{\sqrt{(k^0_1-k^0_2-\omega_3(q,\theta))^2 -\omega^2_4(q,\theta,0{\rm\,or\,}\pi)+i\varepsilon}}.
\label{eq:ibphi}
\end{align}
It is important to determine the singularity of $\cos\theta$, denoted as $\cos\theta_{\rm on} = Z(q)+ i\varepsilon^\prime$, from the above equation.
The infinitesimal imaginary part $\varepsilon^\prime$ could be positive or negative.
The specific expression for $Z(q)$ as a function of $q$ is available, but it is too complicated and not necessary. 
All we need to know is the differential function $\mathrm{d}{Z(q)}/\mathrm{d}{q}$, and the sign of the imaginary part of $\cos\theta_{\rm on}$, which is similar to Eq.~(\ref{eq:costheta}).

Firstly, we consider the sign of the imaginary part of $\cos\theta_{\rm on}$,.
We denote the square of the denominator in Eq.~(\ref{eq:ibphi}) as a function of $\cos\theta$ and $q$ as:
\begin{align}
f_b(\cos\theta,q)\equiv& (k^0_1-k^0_2-\omega_3(q,\theta))^2
  -\omega^2_4(q,\theta,0{\rm\,or\,}\pi).\label{eq:fffbbb}
\end{align}
Then we have
\begin{align}
f_b(Z+i\varepsilon^\prime,q)+i\varepsilon=&f_b(Z,q)+ \left.\frac{\partial f_b}{\partial \cos\theta}\right|_{\cos\theta=Z}i\varepsilon^\prime+i\varepsilon,\label{eq:fexpand}
\end{align}
where
\begin{align}
\left.\frac{\partial f_b}{\partial \cos\theta}\right|_{\cos\theta=Z}=& \frac{2q\omega_4(q,\theta_{\rm on},0{\rm\,or\,}\pi)}{\sin\theta_{\rm on}}(v_{3\perp1}-v_{4\perp1}).
\end{align}
Here, $v_{3\perp1}$ and $v_{4\perp1}$ are the velocities of intermediate particles 3 and 4 perpendicular to the direction of particle 1,
\begin{align}
v_{3\perp1} =& \frac{k_2\sin\theta_{\rm on}}{\omega_3(q,\theta_{\rm on})}, \nonumber \\
v_{4\perp1} =&\left\{
\arraycolsep=4pt\def\arraystretch{1.3}
\begin{array}{ll}
\frac{-k_2\sin(\theta_{\rm on}-\theta_1)}{\omega_4(q,\theta_{\rm on},0)},&  \phi=0 \\
\frac{-k_2\sin(\theta_{\rm on}+\theta_1)}{\omega_4(q,\theta_{\rm on},\pi)},&  \phi=\pi  
\end{array}\right..
\label{eq:vb134b}
\end{align}
Here we take $v_{3\perp1}>0$, and the sign of $v_{4\perp1}$ is undetermined.
%
It is evident that the sign of $\varepsilon^\prime$ in Eq.~(\ref{eq:fexpand}), which is the imaginary part of $\cos\theta_{\rm on}$, is determined by the sign of $\frac{\partial f_b}{\partial \cos\theta}$, so we have:
\begin{align}
\cos\theta_{\rm on}= 
\left\{
\def\arraystretch{1.1}
\begin{array}{ll}
Z_1(q)-i\varepsilon_0, &  v_{3\perp1}>v_{4\perp1} \\
Z_2(q)+i\varepsilon_0, &  0<v_{3\perp1}<v_{4\perp1} 
\end{array}
\right..
\label{eq:costhetabbb}
\end{align}

Then ${\cal I}_{b_1}$ can be expressed as:
\begin{align}
{\cal I}_{b_1} \propto& \int\frac{q^2\mathrm{d}{q}{\cal \tilde{I}}^\prime_b}{M-\omega_1(q)-\omega_2(q)+i\varepsilon}, \label{eq:interIb1ab} \\
{\cal \tilde{I}}^\prime_b =& \int\frac{\mathrm{d}{\cos\theta}}
  {(\cos\theta-Y(q)+i\varepsilon)\sqrt{\cos\theta-\cos\theta_{\rm on}}}. \label{eq:interIb1a}
\end{align}
$Y(q)$ is given in Eq.~(\ref{eq:xyq}).
It is obvious that ${\cal I}_{b_1}$ has a pole at $q_{\rm on}+i\varepsilon$, so it is necessary that ${\cal \tilde{I}}^\prime_b$ has a pole at $q_{\rm on}-i\varepsilon$ to make ${\cal I}_{b_1}$ diverge.
This is quite similar to ${\cal \tilde{I}}^\prime_a$ in the previous section, where there are two different ways to make ${\cal \tilde{I}}^\prime_a$ diverge after integrating with $\cos\theta$: one is that the integral path is pinched between two $\cos\theta$ poles from two denominators, and the other is to have endpoint divergence of $\cos\theta$.
As discussed before, for $\theta_1\neq 0$ and $\pi$, the endpoint divergence does not produce box singularities but triangle singularities. 
So, we only need to consider the case where the integral path of $\cos\theta$ is pinched between two poles, which we still refer to as the general box singularity, and this will be discussed in detail in the first subsection.
In the second subsection, we will discuss the cases of $\theta_1= 0$ and $\pi$ one by one.

\subsection{The general box singularity}
\label{sec:noTSB}

In this subsection, we only consider that the singularity is generated when the integral path of $\cos\theta$ is pinched between two poles.
The pole of the first denominator of Eq.~(\ref{eq:interIb1a}) locates below the real axis of $\cos\theta$, while the imaginary part of the pole of the second denominator can be positive or negative as shown in Eq.~(\ref{eq:costhetabbb}).
%
%
Only the solution with a positive imaginary part in Eq.(\ref{eq:costhetabbb}) is what we need, and the corresponding divergence condition of ${\cal \tilde{I}}^\prime_{b}$ is $0<v_{3\perp1}<v_{4\perp1}$.
%
%
%

\begin{figure}[htbp]
\centering
  \includegraphics[width=0.8\linewidth]{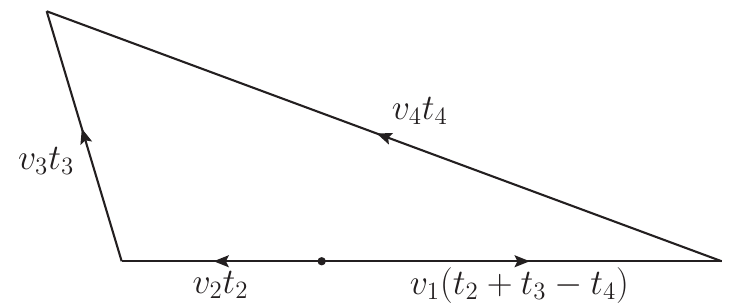}
  \caption{The diagram of the internal particle trajectories in real space-time when box singularity happens in the case of Fig.~\ref{fg:3loopb}.}
  \label{Fig:pattern1B}
\end{figure}

Then for ${\cal \tilde{I}}_b^\prime$, we have:
\begin{align}
  {\cal \tilde{I}}_b^\prime \propto& \int\frac{\mathrm{d}{\cos\theta}}{(\cos\theta-Y(q)+i\varepsilon)\sqrt{\cos\theta-Z_2(q)-i\varepsilon}} \nonumber \\
  \propto&\frac{1}{\sqrt{Y(q)-Z_2(q)-i\varepsilon}}.
\end{align}
Thus, the divergence of the integral ${\cal \tilde{I}}_b^\prime$ comes from the $Y(q)-Z_2(q)-i\varepsilon$ term. 
It means that ${\cal I}_{b_1}$ is divergent when 
\begin{equation}
    Y(q_{\rm on}-i \varepsilon)-Z_2(q_{\rm on}-i \varepsilon)-i\varepsilon^\prime=0. \label{Eq:qonamB}
\end{equation}
The real part of Eq.(\ref{Eq:qonamB}) is satisfied when the particles are on-shell, and for the imaginary part to also be valid, it requires 
\begin{align}
  0 >& \frac{{\rm d}(Y-Z_2)}{{\rm d}q}\Big|_{q\to q_{\rm on}}=\frac{{\rm d}Y}{{\rm d}q}\Big|_{q\to q_{\rm on}}-\frac{{\rm d}Z_2}{{\rm d}q}\Big|_{q\to q_{\rm on}}.
  \label{Eq:imYZB1}
\end{align}
%
We use the definition of $f_b$ in Eq.(\ref{eq:fffbbb}) to calculate $\frac{{\rm d}Z_2}{{\rm d}q}$, that is
\begin{align}
\frac{{\rm d}Z_2}{{\rm d}q}=-\frac{\partial f_b(Z_2, q) / \partial q}{\partial f_b(Z_2, q) / \partial Z_2},
\label{eq:z2q}
\end{align}
where $f_b(Z_2(q),q)=0$.
It is straightforward to obtain:
\begin{align}
\frac{\partial f_b(Z_2,q)}{\partial q}=
2\omega_4(q,\theta_{\rm on},\phi=0{\rm\,or\,}\pi)(v_{3\parallel1}-v_{4\parallel1}),
\label{Eq:dfdq}
\end{align}
where
\begin{align}
v_{3\parallel1} =&\frac{k_1\cos \theta_{\rm on}-q}{\omega_3(q,\theta_{\rm on})}, \nonumber \\
v_{4\parallel1} =&\left\{
\arraycolsep=4pt\def\arraystretch{1.3}
\begin{array}{ll}
\frac{q-k_2\cos(\theta_1-\theta_{\rm on})}{\omega_4(q,\theta_{\rm on},0)},&  \phi=0 \\
\frac{q-k_2\cos(\theta_1+\theta_{\rm on})}{\omega_4(q,\theta_{\rm on},\pi)},&  \phi=\pi  
\end{array}
\right..
\label{eq:v4p1b}
\end{align}
So we can obtain the singularity condition based on Eqs.~(\ref{Eq:imYZB1}, \ref{eq:z2q}) as follows, 
\begin{align}
 0 >& \frac{\omega_3(q_{\rm on},\theta_{\rm on})}{k_1 q_{\rm on}}
  \frac{v_{3\perp1}v_{4\perp1}}{v_{4\perp1}-v_{3\perp1}} \left( \frac{v_{4\parallel1}-v_1}{v_{4\perp1}} -\frac{v_{3\parallel1}-v_1}{v_{3\perp1}} \right),
\label{Eq:imYZB}
\end{align}
which is equivalent to
\begin{equation}
    \frac{v_{4\parallel1}-v_1}{v_{4\perp1}} -\frac{v_{3\parallel1}-v_1}{v_{3\perp1}} < 0.
     \label{Eq:imYZB2}
\end{equation}
Here $v_1=q_{\rm on}/\omega_1(q_{\rm on})$ and the definitions of $v_{3\perp1}$, $v_{4\perp1}$, $v_{3\parallel1}$ and $v_{4\parallel1}$ are given in Eqs.(\ref{eq:vb134b}) and (\ref{eq:v4p1b}) with $q=q_{\rm on}$.
Similar to the case of Fig.~\ref{fg:3loopa}, we can physically explain Eq.(\ref{Eq:imYZB2}).
Assume that the lifetimes of particles 2, 3, and 4 are $t_2$, $t_3$, and $t_4$, respectively.
For particle 3 and particle 4 to collide, we have $v_{3\perp1}t_3=v_{4\perp1}t_4$ and $v_2t_2+v_{3\parallel1}t_3=v_1(t_2+t_3-t_4)+v_{4\parallel1}t_4$.
Since $v_1$ is positive, $v_2$ is negative and $t_2$ is larger than zero, we have:
\begin{align}
  (v_1-v_2)t_2 =&v_{3\parallel1}t_3-v_{4\parallel1}t_4-v_1(t_3-t_4) \nonumber \\
  =& (\frac{v_{3\parallel1}-v_1}{v_{3\perp1}} - \frac{v_{4\parallel1}-v_1}{v_{4\perp1}})v_{3\perp1}t_3 > 0. \label{Eq:pattern1B}
\end{align}
It can be seen that Eq.(\ref{Eq:pattern1B}) is exactly the same as Eq.(\ref{Eq:imYZB2}).
So the physical explanation of Eq.(\ref{Eq:imYZB2}) is that particle 4 can collide with particle 3, and this physical process is vividly illustrated in Fig.~\ref{Fig:pattern1B}.

Therefore, the singularity conditions of ${\cal I}_{b_1}$ are:
\begin{align}
  &v_{4\perp1}>v_{3\perp1}>0,\label{eq:v4gv3111}\\
  &\frac{v_{4\parallel1}-v_1}{v_{4\perp1}} -\frac{v_{3\parallel1}-v_1}{v_{3\perp1}} < 0. \label{Eq:patternb111}
\end{align}
Then the singularity conditions for ${\cal I}_{b_2}$ can be easily obtained by exchanging particle indices, that is,
\begin{align}
  &v_{3\perp2}>v_{4\perp2}>0,\\
  &\frac{v_{3\parallel2}-v_2}{v_{3\perp2}} -\frac{v_{4\parallel2}-v_2}{v_{4\perp2}} < 0. \label{Eq:patternb222}
\end{align}
where the value of $v_2$ is set to be positive.
However, we need to write this condition in the same framework, that is, $v_1$ is positive and $v_2$ is negative.
So we have:
\begin{align}
  &v_{3\perp1}>v_{4\perp1}>0,\label{eq:v4gv3222}\\
  &\frac{v_{4\parallel1}-v_2}{v_{4\perp1}}-\frac{v_{3\parallel1}-v_2}{v_{3\perp1}} < 0. \label{Eq:patternb222b}
\end{align}
%
%
Similarly, there are $v_{3\perp1}t_3=v_{4\perp1}t_4$ and $v_2(t_1+t_4-t_3)+v_{3\parallel1}t_3=v_1t_1+v_{4\parallel1}t_4$ to make particle 3 and 4 collide in real space-time.
Then there are:
\begin{align}
  (v_1-v_2)t_1 =&v_{3\parallel1}t_3-v_{4\parallel1}t_4-v_2(t_3-t_4) \nonumber \\
  =& (\frac{v_{3\parallel1}-v_2}{v_{3\perp1}} - \frac{v_{4\parallel1}-v_2}{v_{4\perp1}})v_{3\perp1}t_3 > 0,
\label{Eq:pattern1B2}
\end{align}
which is the same as Eq.(\ref{Eq:patternb222b}).
Therefore, the physical interpretation of Eq.(\ref{Eq:patternb222b}) is that particles 4 and 3 can actually collide. 
The corresponding physical process is also shown in Fig.~\ref{Fig:pattern1B}, which is the same as the case of $v_{4\perp1} > v_{3\perp1}$.

To sum up, the physical manifestation of singularity is that particle 4 can collide with particle 3 in real space-time.

\subsection{For the case of \texorpdfstring{$\theta_1=0$}{theta1=0} or \texorpdfstring{$\pi$}{pi}}
\label{sec:olineB}

In this subsection, we aim to obtain the singularity condition when $\theta_1=0$ or $\pi$.
At this point, variable $\phi$ no longer appears in the integration, and correspondingly, the energy of particle 4 becomes
\begin{align}
  \omega^{\mp}_4(q,\theta)=&\sqrt{q^2+k^2_2+m_4^2\mp 2qk_2\cos\theta},
\label{eq:omega4}
\end{align}
where $\mp$ correspond to $\theta_1=0$ and $\pi$, respectively.
${\cal I}_{b_1}$ can be simplified as follows,
\begin{align}
  {\cal I}_{b_1} =& \int\frac{2\pi q^2\mathrm{d}{q}{\cal \tilde{I}}^\prime_b}{M-\omega_1(q)-\omega_2(q)+i\varepsilon}, \nonumber \\
  {\cal \tilde{I}}^\prime_b =& \int\frac{\mathrm{d}{\cos\theta}}{k^0_1-\omega_1(q)-\omega_3(q,\theta)+i\varepsilon} \nonumber \\
  &\times \frac1{k^0_1-k^0_2-\omega_3(q,\theta) - \omega^{\mp}_4(q,\theta)+i\varepsilon}.
\label{eq:interIb2}
\end{align}
The pole of the denominator of ${\cal I}_{b_1}$ is $q=q_{\rm on}+i\varepsilon$, so it requires ${\cal \tilde{I}}^\prime_b(q_{\rm on}-i\varepsilon)\to \infty$ to make ${\cal I}_{b_1}$ diverge.

We then analyze the integral ${\cal \tilde{I}}^\prime_b$, whose two denominators each produce a first-order pole.
As mentioned before, there are two ways for ${\cal \tilde{I}}^\prime_b$ to diverge.
The pole of the first denominator of ${\cal \tilde{I}}^\prime_b$ is $\cos\theta=Y(q)-i\varepsilon$, so for the case where the two poles are pinched, we only need the imaginary part of the pole of the second denominator to be positive.
We can obtain the imaginary part of the second denominator at the pole with $\cos\theta = Z_b+i\varepsilon_Z$ is as follows, 
%
\begin{align}
 q\left(\frac{k_1}{\omega_3}\pm\frac{k_2}{\omega^\mp_4}\right)\varepsilon_Z+\varepsilon  .
\label{eq:epsilonz}
\end{align}
It is easy to find that $\varepsilon_Z$ is positive only when $\theta_1=\pi$ and $\frac{k_1}{\omega_3(q,\theta)}<\frac{k_2}{\omega_4(q,\theta)}$ when such term to be zero.
We denote the solution under this condition as $Z_b(q)$, and then there is: 
\begin{align}
  {\cal \tilde{I}}^\prime_b\propto &\int\frac{\mathrm{d}{\cos\theta}}{(\cos\theta-Y(q)+i\varepsilon)(\cos\theta-Z_b(q)-i\varepsilon)} \nonumber \\
  \propto &\frac{1}{Y(q)-Z_b(q)-i\varepsilon}.
\label{eq:interib4}
\end{align}
Although we do not provide the explicit expression for $Z_b(q)$, we still need to confirm the sign of the following equation,
\begin{align}
  &\text{Im}[Y(q_{\rm on}-i\varepsilon^\prime)-Z_b(q_{\rm on}-i\varepsilon^\prime)]-\varepsilon \nonumber \\
 =& -\varepsilon^\prime\left.\frac{\mathrm{d}{\left(Y(q)-Z_b(q)\right)}}{\mathrm{d}q}\right|_{q\to q_{\rm on}}-\varepsilon.
\label{eq:interib8}
\end{align}
Through calculations similar to Eq.(\ref{Eq:imYZB1}), we obtain $\mathrm{d}{Z_b(q)}/\mathrm{d}{q}$ as:
\begin{align}
  \frac{{\rm d}Z_b}{{\rm d}q}=&-\frac{\partial f_b(Z_b,q) / \partial q}{\partial f_b(Z_b,q) / \partial Z_b},
\end{align}
where $f_b(\cos\theta,q)\equiv k^0_1-k^0_2-\omega_3(q,\theta)- \omega^+_4(q,\theta)$ and $f_b(Z_b(q),q)=0$ is satisfied.
Then we have: 
\begin{align}
& \left.\frac{\mathrm{d}{(Y(q)-Z_b(q))}}{\mathrm{d}{q}}\right|_{q\to q_{\rm on}} \nonumber \\
=& \left.\left(-\frac{\frac{\partial \omega_{1}}{\partial q}}{\frac{\partial \omega_{3}}{\partial \cos\theta}}+
\frac{-\frac{\partial \omega_{3}}{\partial q}\frac{\partial \omega^+_{4}}{\partial \cos\theta}
 +\frac{\partial \omega^+_{4}}{\partial q}\frac{\partial \omega_{3}}{\partial \cos\theta}}
{\frac{\partial \omega_{3}}{\partial \cos\theta}\left(\frac{\partial \omega_{3}}{\partial \cos\theta}+\frac{\partial \omega^+_{4}}{\partial \cos\theta}\right)}\right)\right|_{\cos\theta\to Z_b(q),\,q\to q_{\rm on}} \nonumber \\
=&\left.\frac{\omega_{3}}{k_1} \left(\frac{1}{\omega_1} + \frac{k_1+k_2}{\omega^+_4k_1-\omega_3k_2}\right) \right|_{\cos\theta\to Z_b(q),\,q\to q_{\rm on}} >0.
\end{align}
Therefore, Eq.(\ref{eq:interib8}) is always negative, which means that $q_{\rm on}-i\varepsilon$ cannot be a pole of ${\cal \tilde{I}}^\prime_b$.
In other words, the only way for ${\cal I}_{b_1}$ to diverge in this case is for the pole to be $\cos\theta=\pm 1$.
The corresponding physical situation is that all particles move on the same line.
%
%
There are four different cases where $\theta$ and $\theta_1$ are equal to 0 or $\pi$, and each of these will be discussed in the next four subsections.

\subsubsection{\texorpdfstring{$\theta_1=\theta=\pi$}{theta1=theta=pi}}

For $\theta_1=\theta=\pi$, ${\cal \tilde{I}}^\prime_{b}$ can be calculated as:
\begin{align}
{\cal \tilde{I}}^\prime_{b} \propto& \int \frac{\mathrm{d}\cos\theta }{(\cos\theta-(Y(q)-i\varepsilon_Y))(\cos\theta-(Z_b(q)+i\varepsilon_Z))} \nonumber \\
\propto& \frac{\log(-1-(Y(q)-i\varepsilon_Y))-\log(-1-(Z_b(q)+i\varepsilon_Z))}{Y(q)-Z_b(q)-i(\varepsilon_Y+\varepsilon_Z)},
\label{eq:iablinepipib}
\end{align}
where $\varepsilon_Y>0$ and the sign of $\varepsilon_Z$ is undetermined.
%
%
Then with $Y(q_{\text{on}})=-1$ and $Z_b(q_{\text{on}})=-1$, we have:
\begin{align}
Y(q_{\text{on}}-i\varepsilon) =& Y(q_{\text{on}})-i\varepsilon\left.\frac{\mathrm{d} Y(q)}{\mathrm{d}{q}}\right|_{q\to q_{\text{on}}} \nonumber \\
=& -1-\frac{i\varepsilon}{q_{\text{on}}}\left(\frac{q_{\text{on}}/\omega_1(q_{\rm on})}{k_1/k^0_1}+1\right),\label{eq:yyb}\\
Z_b(q_{\text{on}}-i\varepsilon) =& Z_b(q_{\text{on}})-i\varepsilon\left.\frac{\mathrm{d} Z_b(q)}{\mathrm{d}{q}}\right|_{q\to q_{\text{on}}} \nonumber \\
=& -1-\frac{i\varepsilon}{q_{\rm on}}
\left(\frac{k_1}{\omega_3(q_{\rm on},\pi)}-\frac{k_2}{\omega^+_4(q_{\rm on},\pi)}\right)^{-1}
\nonumber\\
&\times\left(\frac{q_{\rm on}+k_1}{\omega_3(q_{\rm on},\pi)}+\frac{q_{\rm on}-k_2}{\omega^+_4(q_{\rm on},\pi)}\right), 
\end{align}
We find that the imaginary part of $-1-(Y(q_{\rm on}-i\varepsilon)-i\varepsilon_Y)$ is always positive, which means that $q_{\rm on}-i\varepsilon$ cannot be the pole of $\log(-1-(Y(q)-i\varepsilon_Y))$.
While for the $\log(-1-(Z_b(q)+i\varepsilon_Z))$ term, we have:
\begin{align}
&-1-(Z_b(q_{\text{on}}-i\varepsilon)+i\varepsilon_Z) = i\varepsilon\left.\frac{\mathrm{d} Z_b(q)}{\mathrm{d}{q}}\right|_{q\to q_{\rm on}}-i\varepsilon_Z \nonumber \\
=& -i\varepsilon_Z+\frac{i\varepsilon}{q_{\rm on}} \left(\frac{k_1}{\omega_3(q_{\rm on},\pi)}-\frac{k_2}{\omega^+_4(q_{\rm on},\pi)}\right)^{-1} \nonumber \\
&\times\left(\frac{q_{\rm on}+k_1}{\omega_3(q_{\rm on},\pi)}+\frac{q_{\rm on}-k_2}{\omega^+_4(q_{\rm on},\pi)}\right) \nonumber \\
=& -i\varepsilon_Z+\frac{i\varepsilon}{q_{\rm on}} \left(\frac{k_2}{\omega^+_4(q_{\rm on},\pi)}-\frac{k_1}{\omega_3(q_{\rm on},\pi)}\right)^{-1}(v_4-v_3).
\label{eq:zzb}
\end{align}
Since the sign of $\varepsilon_Z$ is the same as that of $\frac{k_2}{\omega^+_4(q_{\rm on},\pi)}-\frac{k_1}{\omega_3(q_{\rm on},\pi)}$ as shown in Eq.~(\ref{eq:epsilonz}), we can see that $q_{\rm on}-i\varepsilon$ can be a pole only when $v_4>v_3>0$.
Actually, when $v_4>v_3>0$, $\frac{k_2}{\omega^+_4(q_{\rm on},-1)}>\frac{k_1}{\omega_3(q_{\rm on},-1)}$ will be satisfied which leads to $\varepsilon_Z>0$.
Thus, when $\varepsilon_Z<0$, $q_{\rm on}-i\varepsilon$ cannot be a pole of the two $\log$ terms in $\tilde{\cal I}_b^\prime$.
Through the discussions similar to those in Sec.~\ref{sec:0pia}, we can conclude that $q_{\text{on}}-i\varepsilon$ is not a pole of the denominator of Eq.~(\ref{eq:iablinepipib}).
In short, $q_{\text{on}}-i\varepsilon$ is a pole of ${\cal \tilde{I}}^\prime_{b}$ only when $v_4>v_3>0$.

The complete integral ${\cal I}_{b_1}$ can be expressed as:
\begin{align}
{\cal I}_{b_1} \propto& \int \mathrm{d}q \frac{\log(q-q_{\text{on}}-i\varepsilon_1)-\log(q-q_{\text{on}}+i\varepsilon)}{(q-q_{\text{on}}-i\varepsilon_2)(q-q_{\text{on}}-i\varepsilon_3)}.
\end{align}

\begin{figure}[htbp]
 \centering
\subfigure[]{\label{Fig:pipib} \includegraphics[width=0.93\linewidth]{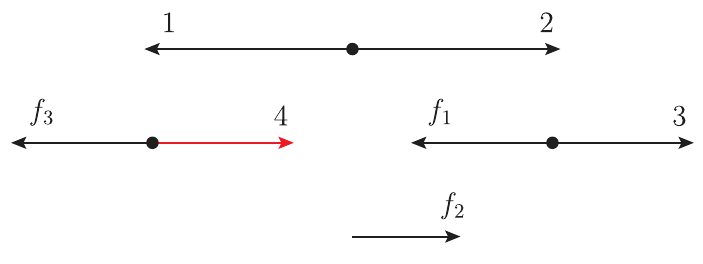}}
 \subfigure[]{\label{Fig:0pib} \includegraphics[width=0.93\linewidth]{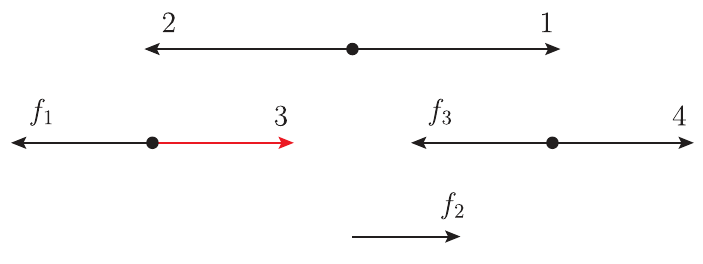}}
 \caption{The directions of each particle in the case of $\theta_1=\theta=\pi$. Black line: the direction is determined; red line: the direction is determined only when singularity occurs.}
\end{figure}

We then turn to the physical condition of this case.
When the singularity occurs, particles $2$, $3$, and $f_2$ move along the $\hat z$-axis, while particles $1$, $f_1$ and $f_3$ move in the opposite direction, as shown in Fig.~\ref{Fig:pipib}.
The singularity condition also includes $v_4>v_3>0$, which indicates that particle 4 is able to catch up with particle 3 and then generate particle $f_2$.
This can be recognized as a classical process.

At last, we consider the singularity condition of ${\cal I}_{b_2}$ by exchanging the indices of particles 1 and 2, 3 and 4, $f_1$ and $f_3$.
Correspondingly, the directions of each particle are given in Fig.~\ref{Fig:0pib}, and the condition is also clear that $v_3>v_4>0$. 
%
%


\subsubsection{\texorpdfstring{$\theta_1=\pi$}{theta1=pi} and \texorpdfstring{$\theta=0$}{theta=0}}

Similar to Eq.(\ref{eq:iablinepipib}), for $\theta_1=\pi$ and $\theta=0$, ${\cal \tilde{I}}^\prime_{b}$ is obtained as:
\begin{align}
{\cal \tilde{I}}^\prime_{b} \propto&
\frac{\log(1-(Y(q)-i\varepsilon_Y))-\log(1-(Z_b(q)+i\varepsilon_Z))}{Y(q)-Z_b(q)-i(\varepsilon_Y+\varepsilon_Z)}.
  \label{eq:iabline0pib}
\end{align}
Based on Eq.(\ref{eq:yyb}), it can be inferred that when $v_3>v_1>0$ and $Y(q_{\rm on})=1$, $q_{\rm on}-i\varepsilon$ can be a pole of $\log(1-(Y(q)-i\varepsilon_Y)$. 
For the $\log(1-(Z_b(q)+i\varepsilon_Z))$ term, the condition for divergence at $q_{\rm on}-i\varepsilon$ is $v_3>v_4>0$ and $Z(q_{\rm on})=1$.
The latter condition happens to be the same as the divergence condition of ${\cal I}_{b_2}$ discussed at the end of the previous subsection. 
The corresponding physical situation is also clear, that is, particle 3 can catch up with particle 1 or 4, and it is the same as shown in Fig.~\ref{Fig:0pib}.
However, if particle 1 decays into particle 4 and $f_3$ before particle 3 catches up with it, and $v_3<v_4$, then such a process cannot be regarded as a classical process. 
As discussed at the end of the previous section, when considering particle lifetimes, the singularity condition merely requires that classical processes can occur without insisting that they must occur. 

In addition, the singularity condition of ${\cal I}_{b_2}$ can also be obtained by exchanging particle indices, i.e. $v_4>v_2>0$ or $v_4>v_3>0$, which is the same as that of ${\cal I}_{b_1}$ with $\theta_1=\theta=\pi$, and its corresponding physics is shown in Fig.~\ref{Fig:pipib}.

\subsubsection{\texorpdfstring{$\theta_1=\theta=0$}{theta1=theta=0}}

\begin{figure}[htbp]
 \centering
 \subfigure[]{\label{Fig:00ba} \includegraphics[width=0.93\linewidth]{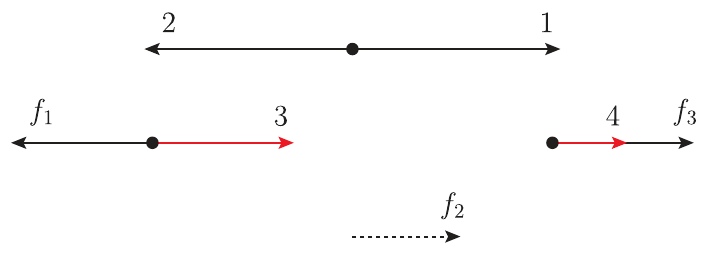}}
 \subfigure[]{\label{Fig:00bb} \includegraphics[width=0.93\linewidth]{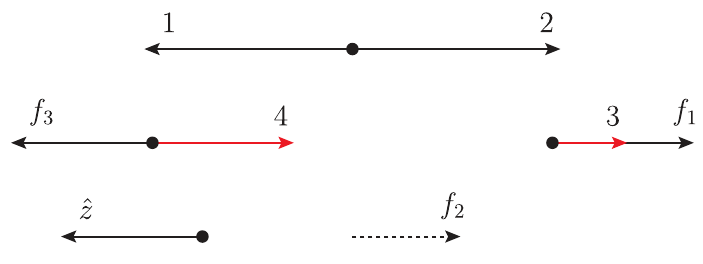}}
 \caption{The directions of each particle in the case of $\theta_1=\theta=0$. Black line: the direction is determined; red line: the direction is determined only when singularity occurs. Black dashed line: the direction is uncertain and is only drawn as an example.}
 \label{Fig:00b}
\end{figure}

%
In this case, we have:
\begin{align}
{\cal \tilde{I}}^\prime_{b} \propto&
\frac{\log(1-(Y(q)-i\varepsilon_Y))-\log(1-(Z_b(q)-i\varepsilon_Z))}{Y(q)-Z_b(q)-i(\varepsilon_Y-\varepsilon_Z)},
\end{align}
where $\varepsilon_Y$ and $\varepsilon_Z$ are positive.
We find that both $\log$ terms could have a pole at $q_{\rm on}-i\varepsilon$, and their corresponding conditions are $v_3>v_1>0$ and $v_3>v_4$, respectively.
It is natural to obtain that the singularity condition of ${\cal I}_{b_1}$ is $v_3>v_1>0$ or $v_3>v_4$, and it corresponds to the situation that particle 3 can catch up with particle 1 or 4, as shown in Fig.~\ref{Fig:00ba}.
Meanwhile, the singularity condition of ${\cal I}_{b_2}$ is very similar, that is $v_4<v_2<0$ or $v_4<v_3$, and the corresponding physics is shown in Fig.~\ref{Fig:00bb}.
It should be noted that the directions of the $\hat z$-axis in Figs.~\ref{Fig:00ba} and \ref{Fig:00bb} are opposite.

In summary, the conditions for box singularity are $v_3>v_1>0$, $v_3>v_4$, or $v_4<v_2<0$, and of course, the condition that all four intermediate particles are on-shell still needs to be met.

\subsubsection{\texorpdfstring{$\theta_1=0$}{theta1=0} and \texorpdfstring{$\theta=\pi$}{theta=pi}}

\begin{figure}[htbp]
 \centering
 \subfigure[]{\label{Fig:pi0ba} \includegraphics[width=0.93\linewidth]{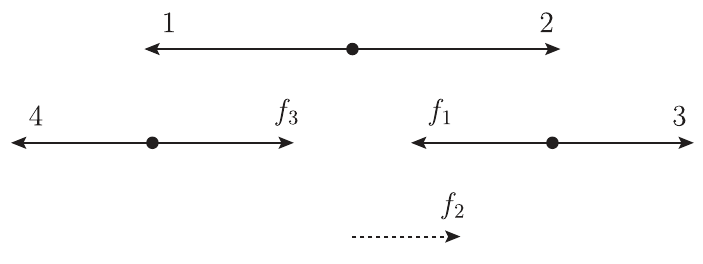}}
 \subfigure[]{\label{Fig:pi0bb} \includegraphics[width=0.93\linewidth]{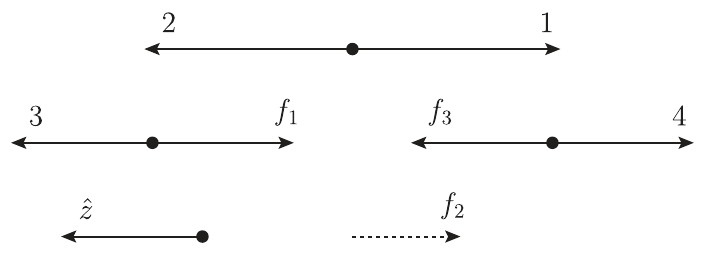}}
 \caption{The directions of each particle in the case of $\theta_1=0$ and $\theta=\pi$. Solid line: the direction is determined; dashed line: the direction is determined only when singularity occurs.}
\label{Fig:pi0b}
\end{figure}

For the last case where $\theta_1=0$ and $\theta=\pi$, there is:
\begin{align}
{\cal \tilde{I}}^\prime_{b} \propto&
\frac{\log(-1-(Y(q)-i\varepsilon_Y))-\log(-1-(Z_b(q)-i\varepsilon_Z))}{Y(q)-Z_b(q)-i(\varepsilon_Y-\varepsilon_Z)},
\label{eq:iabline00b}
\end{align}
where $\varepsilon_Y>0$ and $\varepsilon_Z>0$.
Here, $q_{\rm on}-i\varepsilon$ cannot be the pole of these two $\log$ terms.
The sign of the imaginary part of the denominator in Eq.(\ref{eq:iabline00b}) at $q_{\rm on}-i\varepsilon$ is indeterminate.
Through the same discussion as in Sec.~\ref{sec:0pia}, we can know that $q_{\rm on}-i\varepsilon$ cannot be the pole of ${\cal \tilde{I}}^\prime_{b}$.
Thus, for ${\cal I}_{b_1}$ with integral over $q$, it is not divergent. 
Fig.~\ref{Fig:pi0ba} shows the corresponding physical situation at this point, indicating that the intermediate particles cannot undergo classical collisions.
In addition, it can be concluded that ${\cal I}_{b_2}$ also converges in this case, and its corresponding physics is given in Fig.~\ref{Fig:pi0bb}.

\section{Mass condition formulae for generating singularity}
\label{sec:formula}

In the previous sections, we provided the singularity conditions based on velocity relations in various cases of box diagrams, which can be understood from Coleman-Norton theorem.
However, the conditions expressed in terms of velocity relations cannot be directly applied to determine whether a singularity exists in a particular box diagram.
The best way is to express the conditions as relations between masses, which can effectively determine whether there is Landau singularity in a given box diagram.

For the triangle singularity, the authors provide a formula that relates the masses of the involved particles in Ref.~\cite{Guo:2019twa}.
We do the same here.
For a triangle loop diagram, if all the intermediate particles are on-shell and all move in the same straight line, then there is a mass relation:
\begin{align}
4M^2m_3^2 = & (m^2_1-m^2_2+m^2_{f_1}-m_{f_2}^2)^2 \nonumber \\
& -\left(\zeta(M,m_1,m_2)- \zeta(M,m_{f_1},m_{f_2})\right)^2.
\label{eq:trimass1}
\end{align}
Here we define $\zeta(a,b,c)=\sqrt{\lambda(a^2,b^2,c^2)}$ for simplicity.
In addition, some conditions need to be met to ensure that all intermediate particles can be on-shell, such as $M>m_1+m_2$, $m_2>m_{f_1}+m_3$, and $m_{f_2}>m_1+m_3$, which we refer to as the on-shell condition.
Although these conditions are necessary for the singularity, they are not the focus of our work, and they are just obvious and simple comparisons of particle masses. 
So in the subsequent discussions, we assume that these on-shell conditions are valid and will not provide specific details.
In Sec.~\ref{sec:TS}, we have derived that the condition for triangle singularity is $v_1<v_3$, so we have:
\begin{align}
\frac{\zeta(M,m_{f_1},m_{f_2})}{m^2_i-m^2_{f_1}+m^2_{f_2}} &> \frac{\zeta(M,m_1,m_2)}{m^2_i-m^2_2+m^2_1}.
\label{eq:trimass2}
\end{align}
According to Eqs.(\ref{eq:trimass1}) and (\ref{eq:trimass2}), it is straightforward to determine whether a given triangle diagram has singularities. 

We then proceed to derive similar mass formulae as needed to generate singularities in the box diagrams.
Since the phase space of the three-body final state is two-dimensional, which means there are two independent variables, we first aim to find the curves in the Dalitz plot that satisfy the general box singularity condition.
While for the case that all particles move on the same line, there would be an equation similar to Eq.(\ref{eq:trimass1}).
We then discuss these two situations separately.

For the general box singularity condition of Fig.~\ref{fg:3loopa}, we choose invariant masses $m_{f_1f_2}$ and $m_{f_2f_3}$ as free variables, then it is straightforwardly to obtain $E_{1}$, $E_{2}$, $E_{f_1}$ and $E_{f_3}$ which are the energies of particle 1, 2, $f_1$ and $f_3$, and 
$q=q_1=q_2$, $q_{f_1}$ and $q_{f_3}$ are the values of three momentum of the particle $1$($2$), $f_1$ and $f_3$, respectively.
Then we can obtain:
\begin{align}
0 =& (M-E_{f_1}-E_{f_3})^2-m^2_{f_2} \nonumber \\
& -(q_{3\parallel 1}-q_{4\parallel 1})^2-(q_{3\perp 1}-q_{4\perp 1})^2,
\label{eq:invariantmassa}
\end{align}
where
\begin{align}
q_{3\parallel 1}=& -q-q_{f_1\parallel 1}, \nonumber \\
q_{3\perp 1}=& \sqrt{q_{f_1}^2-q^2_{f_1\parallel 1}}, \nonumber \\
q_{4\parallel 1}=& q_{f_3\parallel 1}-q, \nonumber \\
q_{4\perp 1}=& -\sqrt{q_{f_3}^2-q^2_{f_4\parallel 1}}, \nonumber \\
E_{f_1/2}=& \frac{m_2^2+m_{f_1}^2-m_{3}^2}{2m_2}, \nonumber \\
E_{f_3/1}=& \frac{m_{1}^2+m_{f_3}^2-m_{4}^2}{2m_{1}}, \nonumber \\
q_{f_1\parallel 1}=& \frac{E_2E_{f_1}-m_2E_{f_1/2}}{q}, \nonumber \\
q_{f_3\parallel 1}=& \frac{E_{f_3}E_{f_3}-m_{1}E_{f_3/1}}{q}.
\label{eq:qf3parallel1a}
\end{align}
The $a/b$ in the subscript represents the variable of particle a in the rest frame of particle b.
If not otherwise specified, it is in the rest frame of the initial particle.
The on-shell conditions are still required.
Eq.(\ref{eq:conditionlast}) can be written as:
\begin{align}
0<& \frac{q_{3\parallel 1}E_1-q(E_2-E_{f_1})}{q_{3\perp 1}E_1}-\frac{q_{4\parallel 1}E_1-q(E_{f_3}-E_{1})}{q_{4\perp 1}E_1}.
\label{eq:invariantmassb}
\end{align}
Note that according to the definitions of $q_{3\perp 1}$ and $q_{4\perp 1}$, as part of Eq.(\ref{eq:conditionlast}), $q_{3\perp 1}>0>q_{4\perp 1}$ is naturally satisfied.
In this way, we can use Eqs.(\ref{eq:invariantmassa}) and (\ref{eq:invariantmassb}) to confirm whether any point on the Dalitz plot satisfies the singularity condition.

There is a similar equation for the general box singularity condition for Fig.~\ref{fg:3loopb}.
Due to the reversal of particle 4, Eq.(\ref{eq:invariantmassa}) becomes
\begin{align}
0=&(M-E_{f_1}-E_{f_3})^2-m^2_{f_2} \nonumber\\
&-(q_{3\parallel 1}+q_{4\parallel 1})^2-(q_{3\perp 1}+q_{4\perp 1})^2, 
\label{eq:invariantmassba}
\end{align}
where $q_{4\parallel 1}$ and $q_{4\perp 1}$ are redefined as:
\begin{align}
q_{4\parallel 1}=& q-q_{f_3\parallel 1}, \nonumber \\
q_{4\perp 1}=& \sqrt{q_{f_3}^2-q^2_{f_4\parallel 1}}. 
\label{eq:qf3parallel1b}
\end{align}
It can be seen that the momenta of particles 3 and 4 are consistent in the direction perpendicular to the motion of particle 1.
Additionally, there are certain differences in the on-shell conditions, and they are still required.
Then we have:
\begin{align}
0<&\frac{q_{3\parallel 1}E_1-q(E_2-E_{f_1})}{q_{3\perp 1}E_1}-\frac{q_{4\parallel 1}E_1-q(E_{1}-E_{f_3})}{q_{4\perp 1}E_1},
\label{eq:invariantmassbb}
\end{align}
when $q_{4\perp 1}(E_2-E_{f_1})>q_{3\perp 1}(E_1-E_{f_3})$, and 
\begin{align}
0<&\frac{q_{3\parallel 1}E_2-q(E_2-E_{f_1})}{q_{3\perp 1}E_2}-\frac{q_{4\parallel 1}E_2-q(E_{1}-E_{f_3})}{q_{4\perp 1}E_2},
\label{eq:invariantmassbb2}
\end{align}
when $q_{4\perp 1}(E_2-E_{f_1})<q_{3\perp 1}(E_1-E_{f_3})$, which are Eqs.(\ref{Eq:patternb111}) and (\ref{Eq:patternb222b}), respectively.

Then we turn to the case where all involved particles move on the same line, and it is obvious that $q_{3\perp 1}=q_{4\perp 1}=0$.
So the values of variables $m_{f_1f_2}$ and $m_{f_2f_3}$ can be determined from $q^2_{f_3}=q^2_{f_3\parallel1}$ and $q^2_{f_1}=q^2_{f_1\parallel1}$.
Note that the definition of $q_{f_3\parallel1}$ is different in the cases of Figs.~\ref{fg:3loopa} and \ref{fg:3loopb}. 
We can also check whether the obtained values of $m_{f_1f_2}$ and $m_{f_2f_3}$ satisfy $m_{f_1f_2}=\sqrt{(E_{2\to 23}+E_3)^2-(q_{f_2\to 12}+q_{f_3})^2}$ or $m_{f_1f_2}=\sqrt{(E_{2\to 23}+E_3)^2-(q_{f_2\to 23}-q_{f_3})^2}$, which correspond to the opposite or same directions of particle $f_1$ and particle $f_2$.
Here $q_{f_2\to 23}=\zeta(m_{f_2f_3},m_{f_2},m_{f_3})/(2m_{f_2f_3})$ and $E_{f_2\to 23}=\sqrt{m_{f_2}^2+q^2_{f_2\to 23}}$.
However, so far, these conditions only ensure that all particles are on-shell and move on the same line.
We still need to combine them with the previously obtained velocity conditions.
%
For the case of Fig.~\ref{fg:3loopa}, we obtain the corresponding conditions $v_4>v_1$ or $v_3>v_1$, which means that one of the following two conditions needs to be met:
\begin{align}
\frac{q_{3\parallel 1}}{E_2-E_{f_1}} >& \frac{q}{E_1}, \nonumber \\
\frac{q_{4\parallel 1}}{E_{f_1}-E_{1}} >& \frac{q}{E_1}.
\label{eq:invariantmassac}
\end{align}
Note that once the values of $m_{f_1f_2}$ and $m_{f_2f_3}$ are determined, the momenta of all particles are also determined, so the above conditions are easy to verify.
For Fig.~\ref{fg:3loopb}, we have derived the condition that $v_4<v_2<0$, $v_3>v_1>0$, or $v_3>v_4$, which means that we require one of the following three conditions:
\begin{align}
\frac{q_{4\parallel 1}}{E_{f_1}-E_{1}}<& -\frac{q}{E_2}, \nonumber \\
\frac{q_{3\parallel 1}}{E_2-E_{f_1}}>& \frac{q}{E_1}, \nonumber \\
\frac{q_{3\parallel 1}}{E_2-E_{f_1}}>& \frac{q_{4\parallel 1}}{E_{f_1}-E_{1}}.
\label{eq:invariantmassacdd}
\end{align}

According to our previous discussion, we have provided a systematic method to determine whether the box diagram with fixed internal and external particles exhibits box singularity. 
However, it is still necessary to examine each point in the Dalitz plot that corresponds to $m_{f_1f_2}$ and $m_{f_2f_3}$.
Actually, in principle, in the phase space, $m_{f_1f_2}$ and $m_{f_2f_3}$ have a certain range.
When $m_{f_2f_3}$ takes a certain value between $m_{f_2}+m_{f_3}$ and $M-m_{f_1}$, the range of $m^2_{f_1f_2}$ is $[(E_{2\to 23}+E_3)^2-(q_{f_2\to 12}+q_{f_3})^2, (E_{2\to 23}+E_3)^2-(q_{f_2\to 23}-q_{f_3})^2]$~\cite{ParticleDataGroup:2022pth}.
For example, according to Eqs.(\ref{eq:invariantmassa}) and (\ref{eq:invariantmassb}), we can obtain the conditions for determining the existence of a singularity in pure mass relation.
However, it is rather complicated by using two invariant mass variables. 
Actually, through introducing two angle variables, we finally find the pure mass condition to check the existence of singularity.
For example, for Fig.~\ref{fg:3loopa}, the condition is 
\begin{align}
&F^{-}\le 4m_1^2m_2^2(m_3^2+m_4^2-m_{f_2}^2)\le F^{+},
\label{eq:mass11}
\end{align}
where
\begin{align}
F^{\pm}\equiv&(M^2-m_1^2-m_2^2) \times\nonumber\\
&\,\,\,\left[(m_2^2-m_{f_1}^2+m_3^2)|m_1^2-m_{f_3}^2+m_4^2|\right. \nonumber\\
&\,\,\,\,\,\,\left.
\pm\zeta(m_{f_3},m_4,m_1)\zeta(m_2,m_3,m_{f_1}) \right]\nonumber\\
&+\zeta(M,m_1,m_2)\times\nonumber\\
&\,\,\,\left[(m_2^2-m_{f_1}^2+m_3^2)\zeta(m_{f_3},m_4,m_1)\right. \nonumber\\
&\,\,\,\,\,\,\left.\pm|m_1^2-m_{f_3}^2+m_4^2|\zeta(m_2,m_3,m_{f_1})\right].
\end{align}
The related derivation of the above equation is shown in Appendix~\ref{app:mass pure}, so as an even more complicated case for Fig.~\ref{fg:3loopb}.
%

\section{Summary}
\label{sec:sum}


In this article, we delve into the issue of box singularities in three-body decay processes with box diagrams. 
We conduct a thorough analysis to identify the various kinematic conditions that may give rise to singularities.
We conduct a thorough analysis to identify the various kinematic conditions that may give rise to singularities. 
From a physical standpoint, we establish two primary criteria that must be satisfied: first, all four internal particles must be on-shell; and second, the conditions must allow for classical collisions, in accordance with the Coleman-Norton theorem.

We emphasize that in scenarios where particles do not move in a straight line, specific velocity constraints are applicable. For Fig.~\ref{fg:3loopa}, the velocity condition is given by Eq.(\ref{eq:conditionlast}), and for Fig.~\ref{fg:baba} and \ref{fg:babb}, the conditions are Eqs.(\ref{eq:v4gv3111},\ref{Eq:patternb111}) and Eqs.(\ref{eq:v4gv3222}, \ref{Eq:patternb222b}), respectively.
Conversely, when all particles move along a straight path, the conditions simplify. For Fig.~\ref{fg:3loopa}, the condition is simply ${\rm Max}(v_3, v_4)>v_1$. For Fig.~\ref{fg:3loopb}, at least one of the following three conditions must be satisfied: $v_3>v_1>0$, $v_3>v_4$ and $v_4<v_2<0$.  
Additionally, we provide a pure mass condition for the box singularity, such as Eq.(\ref{eq:mass11}) for  Fig.~\ref{fg:3loopa}.

Based on the outlined kinematic criteria, we present a straightforward method to evaluate the existence of singularities for any given box diagram. 
Furthermore, we provide a formula that can assess the potential existence of singularities based solely on the masses of the involved particles.
This work enhances our understanding of the underlying structures in particle decay processes and offers a clear framework for analyzing potential singular behavior.

\section*{Acknowledgments}

We thank Hao-Jie Jing for the useful discussions.

This work is supported by the National Natural Science Foundation of China under Grant Nos. 12221005, 12175239,
and by the Chinese Academy of Sciences under Grant No. YSBR-101.

\begin{appendix}

\section{The derivation of Eq.(\ref{eq:interb})}
\label{app:bform}

The complete expression of ${\cal I}_3$ is given in Eq.(\ref{eq:3loop}).
Whether we take the poles with positive or negative imaginary parts, there are always two residues contributing to the divergence, while the other two are convergent.
Here we choose the poles above the real axis, and after applying the residue theorem, the two divergent residues are:
\begin{align}
  {\rm Res}\left(I_3, M-\omega_2+i\varepsilon\right)  =& \frac1{(l_1+l_0) (l_2+l_0) (l_3+l_0)} \nonumber \\
  & \times \frac1{(l_4+l_0) l_0 \tilde{a} \tilde{b}}, \nonumber \\
  {\rm Res}\left(I_3, k_1^0-\omega_3+i\varepsilon\right) =& \frac1{-l_0l_1l_2l_3l_4\tilde{c}\tilde{d}},
\end{align}
where $l_0=M-k_1^0-\omega_2+\omega_3$, $l_1=k_1^0+\omega_1-\omega_3$, $l_2=k_1^0-M-\omega_2-\omega_3+i\varepsilon$, $l_3=-2\omega_3+i\varepsilon$ and $l_4=k_1^0-k_2^0-\omega_3+\omega_4$.
%
%
The expressions of $\tilde{a}$, $\tilde{b}$, $\tilde{c}$ and $\tilde{d}$  are given in Eq.(\ref{eq:abcd}), and it should be noted that there exists the relation $\tilde{a}-\tilde{c}=\tilde{b}-\tilde{d}=l_0$.

Then, the sum of the two residues can be simplified to:
\begin{widetext}
\begin{align}
  & {\rm Res}\left(I_3, M-\omega_2+i\varepsilon\right) + {\rm Res}\left(I_3, k_1^0-\omega_3+i\varepsilon\right)
  \nonumber\\
   =& \frac1{l_0} \left[ \frac1{(l_1+l_0)(l_2+l_0)(l_3+l_0)(l_4+l_0)(\tilde{c}+l_0)(\tilde{d}+l_0)} -\frac1{l_1l_2l_3l_4\tilde{c}\tilde{d}} \right] \nonumber \\
  =& - \frac{(l_1+l_0)(l_2+l_0)(l_3+l_0)(l_4+l_0)(\tilde{c}+l_0)(\tilde{d}+l_0)-l_1l_2l_3l_4\tilde{c}\tilde{d}}{l_0 \tilde{c}\tilde{d}(\tilde{c}+l_0)(\tilde{d}+l_0) \prod_{i=1}^4 l_i \prod_{i=1}^4 (l_i+l_0)} \nonumber \\
  =& - \frac{\tilde{c}\tilde{d}[(l_1+l_0)(l_2+l_0)(l_3+l_0)(l_4+l_0) - l_1l_2l_3l_4] + (l_1+l_0)(l_2+l_0)(l_3+l_0)(l_4+l_0)(\tilde{c}+\tilde{d}+l_0)l_0}{l_0 \tilde{c}\tilde{d}(\tilde{c}+l_0)(\tilde{d}+l_0) \prod_{i=1}^4 l_i \prod_{i=1}^4 (l_i+l_0)} \nonumber \\
  =& \frac{-(\tilde{a}+\tilde{d})}{\tilde{a}\tilde{b}\tilde{c}\tilde{d} \prod_{i=1}^4 l_i} - \frac{(l_1+l_0)(l_2+l_0)(l_3+l_0)(l_4+l_0) - l_1l_2l_3l_4}{\tilde{a}\tilde{b}l_0 \prod_{i=1}^4 l_i \prod_{i=1}^4 (l_i+l_0)}. \label{Eq:Res23}
\end{align}
\end{widetext}
As we claimed before, only $\tilde{a}$, $\tilde{b}$, $\tilde{c}$ and $\tilde{d}$ can be zero, and other terms always have finite values.
The degree of divergence of the first term in Eq.(\ref{Eq:Res23}) is greater than that of the second term, telling that the degree of divergence of ${\cal I}_3$ rely on the first term, and all information about the poles is included in this term.
Therefore, when discussing the singularity of ${\cal I}_3$, we can simply consider ${\cal I}_b$ by taking only the first term in Eq.(\ref{Eq:Res23}) with $l_i$ neglected, that is,
\begin{equation}
  {\cal I}_b = \int \mathrm{d}^3{\vec q} \frac{\tilde{a}+\tilde{d}}{\tilde{a}\tilde{b}\tilde{c}\tilde{d}}.
\end{equation}

\section{The derivation of the solutions for \texorpdfstring{$\cos(\theta_1\mp\theta)=Y-i\varepsilon$}{cos=Y-}}
\label{app:Y0pm}

From $\cos(\theta_1-\theta)=Y-i\varepsilon$ with $\phi=0$ and $\theta_1\ge \theta$, we have:
\begin{align}
  \cos\theta =& \cos(\theta_1-\theta)\cos\theta_1+\sin(\theta_1-\theta)\sin\theta_1 \nonumber\\
  =& (Y-i\varepsilon)\cos\theta_1+\sqrt{1-(Y-i\varepsilon)^2}\sin\theta_1\nonumber\\
  =& Y\cos\theta_1+\sqrt{1-Y^2}\sin\theta_1 \nonumber\\
    & +i\varepsilon(-\cos\theta_1+\frac{Y}{\sqrt{1-Y^2}}\sin\theta_1) \nonumber\\
  =& Y\cos\theta_1+\sqrt{1-Y^2}\sin\theta_1+\frac{i\varepsilon\sin\theta}{\sin(\theta_1-\theta)} \nonumber \\
  =& Y_0^+ + i\varepsilon^\prime.
\end{align}
For $\phi=0$ and $\theta_1\le \theta$, it is:
\begin{align}
  \cos\theta =& \cos(\theta_1-\theta)\cos\theta_1+\sin(\theta_1-\theta)\sin\theta_1 \nonumber\\
  =& (Y-i\varepsilon)\cos\theta_1-\sqrt{1-(Y-i\varepsilon)^2}\sin\theta_1\nonumber\\
  =& Y\cos\theta_1-\sqrt{1-Y^2}\sin\theta_1\nonumber\\
    & +i\varepsilon(-\cos\theta_1-\frac{Y}{\sqrt{1-Y^2}}\sin\theta_1)\nonumber\\
  =& Y\cos\theta_1-\sqrt{1-Y^2}\sin\theta_1+\frac{i\varepsilon\sin\theta}{\sin(\theta_1-\theta)} \nonumber \\
  =& Y_0^- - i\varepsilon^\prime.
\end{align}
%

The derivation in the case of $\phi=\pi$ is very similar to that of $\phi=0$. 
When $\theta_1+\theta\le\pi$, we have:
\begin{align}
   \cos\theta =& \cos(\theta_1+\theta)\cos\theta_1+\sin(\theta_1+\theta)\sin\theta_1 \nonumber\\
  =&(Y-i\varepsilon)\cos\theta_1+\sqrt{1-(Y-i\varepsilon)^2}\sin\theta_1\nonumber\\
  =&Y\cos\theta_1+\sqrt{1-Y^2}\sin\theta_1\nonumber\\
    &+i\varepsilon(-\cos\theta_1+\frac{Y}{\sqrt{1-Y^2}}\sin\theta_1)\nonumber\\
  =&Y\cos\theta_1+\sqrt{1-Y^2}\sin\theta_1-\frac{i\varepsilon\sin\theta}{\sin(\theta_1+\theta)} \nonumber \\
  =& Y_0^+ - i\varepsilon^\prime.
\end{align}
For $\theta_1+\theta\ge\pi$, it is:
\begin{align}
  \cos\theta =& \cos(\theta_1+\theta)\cos\theta_1+\sin(\theta_1+\theta)\sin\theta_1 \nonumber\\
  =& (Y-i\varepsilon)\cos\theta_1-\sqrt{1-(Y-i\varepsilon)^2}\sin\theta_1\nonumber\\
  =& Y\cos\theta_1-\sqrt{1-Y^2}\sin\theta_1\nonumber\\
    & +i\varepsilon(-\cos\theta_1-\frac{Y}{\sqrt{1-Y^2}}\sin\theta_1)\nonumber\\
  =& Y\cos\theta_1-\sqrt{1-Y^2}\sin\theta_1-\frac{i\varepsilon\sin\theta}{\sin(\theta_1+\theta)} \nonumber \\
  =& Y_0^- + i\varepsilon^\prime.
\end{align}
%

Combining the above four situations naturally yields Eq.(\ref{eq:costheta}).

In addition, when $\cos\theta$ is at the boundary, there are several particular solutions because the coefficient of the first order of $\varepsilon$ is vanished. 
Then, after expanding to the first nonzero term, we will have, 
\begin{align}
\cos\theta = 
\left\{ 
\arraycolsep=4pt\def\arraystretch{1.3}
\begin{array}{ll}
1-i\varepsilon^3 \frac{\cos\theta_1}{2\sin^{\frac{3}{2}}\theta_1},&  \phi=0\, \&\, \theta=0 \\
\cos\theta_1\pm i\sqrt{\varepsilon}\sin\theta_1,&  \phi=0\, \&\, \theta=\theta_1 \\
-1-i\varepsilon^3 \frac{\cos\theta_1}{2\sin^{\frac{3}{2}}\theta_1},&  \phi=0\, \&\, \theta=\pi \\
1-i\varepsilon^3 \frac{\cos\theta_1}{2\sin^{\frac{3}{2}}\theta_1},&  \phi=\pi\, \&\, \theta=0 \\
-\cos\theta_1\pm i\sqrt{\varepsilon}\sin\theta_1,&  \phi=\pi\, \&\, \theta=\pi-\theta_1 \\
-1-i\varepsilon^3 \frac{\cos\theta_1}{2\sin^{\frac{3}{2}}\theta_1},&  \phi=\pi\, \&\, \theta=\pi 
\end{array}
\right. .
\label{eq:costheta0}
\end{align}
These solutions all correspond to triangle singularity, not box singularity.
For example, if singularity happens with $\theta=0$, it means that in the rest frame of the initial particle, particle $4$ and particle $1$ move in the same direction, while particle $3$ moves in other direction since at this point $\theta_1\neq 0$ or $\pi$. 
In this case, the singularity of $\cos\theta$ happens at the boundary of the integral path.
It should be pointed out that such singularity will occur simultaneously at $\phi=0$ and $\pi$.
Then the box singularity requires both denominators of ${\cal I}^\prime_a$ to be zero.
However, the singularities of the second denominator of ${\cal I}^\prime_a$ in $\phi=0$ and $\pi$ cases have opposite signs.
Therefore, the divergence of the sum of these two singularities will decrease.
In other words, ${\cal I}^\prime_a$ is still divergent, but its divergence is lower than that of the box singularity, which makes the divergence of ${\cal I}_a$ the same as that of the triangle singularity.
Situations are similar for $\theta=\theta_1$ and $\theta=\pi- \theta_1$.
%

\section{The derivation for Eq.(\ref{eq:condition})}
\label{app:condition}

From Eq.(\ref{eq:xyq}), we can get:
\begin{align}
 \frac{{\rm d}X}{{\rm d}q}\Big|_{q\to q_{\rm on}} =&\frac{k_2^0}{k_2\omega_1(q_{\rm on})}-\frac{\cos\theta}{q_{\rm on}}, \nonumber \\
 \frac{{\rm d}Y}{{\rm d}q}\Big|_{q\to q_{\rm on}} =&\frac{k_1^0}{k_1\omega_1(q_{\rm on})}-\frac{Y(q_{\rm on})}{q_{\rm on}},
 \label{eq:dxdy}
\end{align}
and from Eqs.(\ref{eq:Y0Z0pm}) and (\ref{eq:Y0}), we have:
\begin{align}
\frac{{\rm d}Y_0}{{\rm d}Y} = 
\left\{
\arraycolsep=4pt\def\arraystretch{1.3}
\begin{array}{ll}
\cos\theta_1 - \frac{Y}{\sqrt{1-Y^2}} \sin\theta_1,&  \phi=0  \\
\cos\theta_1 + \frac{Y}{\sqrt{1-Y^2}}\sin\theta_1,&  \phi=\pi  
\end{array}
\right. .
\label{eq:dy0dy}
\end{align}
We use $\omega_i$ to represent $\omega_i(q_{\rm on})$ for convenience, which is the on-shell energy of the $i$-th intermediate particle.

For the case of $\phi=0$, we have $Y=\cos(\theta_1-\theta)$ and $\theta_1\ge \theta$. 
Then, Eq.(\ref{eq:imqonam}) becomes:
\begin{align}
 & \frac{{\rm d}X}{{\rm d}q}\Big|_{q\to q_{\rm on}} - \frac{{\rm d}Y_0}{{\rm d}Y}\frac{{\rm d}Y}{{\rm d}q}\Big|_{q\to q_{\rm on}}  \nonumber \\
 =&\left(\frac{k_2^0}{k_2\omega_1}-\frac{\cos\theta}{q_{\rm on}}\right)
 +\frac{\sin\theta}{\sin(\theta_1-\theta)}\left(\frac{k_1^0}{k_1\omega_1}-\frac{\cos(\theta_1-\theta)}{q_{\rm on}}\right)  \nonumber \\
  =& \frac{\sin\theta}{q_{\rm on}} \left(\frac{\omega_1+\omega_4}{k_2\sin\theta}\frac{q_{\rm on}}{\omega_1}
   - \frac{\cos\theta}{\sin\theta} + \frac{\omega_1+\omega_3}{k_1\sin(\theta_1-\theta)}\frac{q_{\rm on}}{\omega_1} \right.  \nonumber\\
  &\left. - \frac{\cos(\theta_1-\theta)}{\sin(\theta_1-\theta)} \right) \nonumber \\
  =& \frac{\sin\theta}{q_{\rm on}} \left[ \frac{\omega_4 }{k_2\sin\theta} \left( \frac{q_{\rm on}}{\omega_1}- \frac{k_2\cos\theta-q_{\rm on}}{\omega_4}\right) \right. \nonumber\\ 
  &\left. + \frac{\omega_3 }{k_1\sin(\theta_1-\theta)} \left(\frac{q_{\rm on}}{\omega_1}
  - \frac{k_1\cos(\theta_1-\theta)-q_{\rm on}}{\omega_3}\right) \right] \nonumber \\
  =& \frac{\sin\theta}{q_{\rm on}} \left[ -\frac{\omega_4}{p_{4\perp1}} \left(\frac{q_{\rm on}}{\omega_1}-\frac{p_{4\parallel1}}{\omega_4}\right)
   + \frac{\omega_3}{p_{3\perp1}} \left(\frac{q_{\rm on}}{\omega_1}-\frac{p_{3\parallel1}}{\omega_3}\right) \right] \nonumber \\
  =& \frac{\sin\theta}{q_{\rm on}} \left(-\frac{v_1-v_{4\parallel1}}{v_{4\perp1}} +\frac{v_1-v_{3\parallel1}}{v_{3\perp1}}\right) < 0.
  \label{Eq:imYZ1}
\end{align}
The variables $v_1$, $v_{3\perp1}$, $v_{4\perp1}$, $v_{3\parallel1}$ and $v_{4\parallel1}$ are defined in Eqs.(\ref{eq:v134}) and (\ref{eq:v4p1}).

For $\phi=\pi$, there are $Y=\cos(\theta_1+\theta)$ and $\theta_1+ \theta\ge \pi$.
Then from Eq.(\ref{eq:imqonam}), we have:
\begin{align}
  & \frac{{\rm d}X}{{\rm d}q}\Big|_{q\to q_{\rm on}} - \frac{{\rm d}Y_0}{{\rm d}Y}\frac{{\rm d}Y}{{\rm d}q}\Big|_{q\to q_{\rm on}} \nonumber\\
 =& \left(\frac{k_2^0}{k_2\omega_1}-\frac{\cos\theta}{q_{\rm on}}\right)
 -\frac{\sin\theta}{\sin(\theta_1+\theta)}\left(\frac{k_1^0}{k_1\omega_1}-\frac{\cos(\theta_1+\theta)}{q_{\rm on}}\right) \nonumber \\
  =& \frac{\sin\theta}{q_{\rm on}} \left[ \frac{\omega_4 }{k_2\sin\theta }\left( \frac{q_{\rm on}}{\omega_1}- \frac{k_2\cos\theta-q_{\rm on}}{\omega_4}\right) \right. \nonumber\\
  &\left. + \frac{\omega_3 }{k_1\sin(\theta_1+\theta-\pi)} \left(\frac{q_{\rm on}}{\omega_1}
  - \frac{k_1\cos(\theta_1+\theta)-q_{\rm on}}{\omega_3}\right) \right] \nonumber \\
  =& \frac{\sin\theta}{q_{\rm on}} \left[ -\frac{\omega_4}{p_{4\perp1}} \left(\frac{q_{\rm on}}{\omega_1}-\frac{p_{4\parallel1}}{\omega_4}\right)
   + \frac{\omega_3}{p_{3\perp1}} \left(\frac{q_{\rm on}}{\omega_1}-\frac{p_{3\parallel1}}{\omega_3}\right) \right] \nonumber \\
  =& \frac{\sin\theta}{q_{\rm on}} \left( -\frac{v_1-v_{4\parallel1}}{v_{4\perp1}} +\frac{v_1-v_{3\parallel1}}{v_{3\perp1}}\right) < 0.
   \label{Eq:imYZ2}
\end{align}
It can be seen that Eqs.(\ref{Eq:imYZ1}) and (\ref{Eq:imYZ2}) are the same.
Since $\sin\theta$ and $q_{\rm on}$ are both positive, we ultimately obtain:
\begin{equation}
  - \frac{v_1-v_{4\parallel1}}{v_{4\perp1}} +\frac{v_1-v_{3\parallel1}}{v_{3\perp1}} < 0.
\end{equation}

In addition, when $\theta_1=\pi$, we have:
\begin{align}
 & \frac{{\rm d}X}{{\rm d}q}\Big|_{q\to q_{\rm on}} - \frac{{\rm d}Y_0}{{\rm d}Y}\frac{{\rm d}Y}{{\rm d}q}\Big|_{q\to q_{\rm on}} \nonumber \\
 =& \left(\frac{k_2^0}{k_2 \omega_1}-\frac{\cos\theta}{q_{\rm on}} \right) + \left( \frac{k_1^0}{k_1\omega_1} + \frac{\cos\theta}{q_{\rm on}} \right) \nonumber \\
 =& \frac{1}{\omega_1} \left( \frac{k_2^0}{k_2} + \frac{k_1^0}{k_1} \right).
  \label{Eq:imYZ0}
\end{align}
It is obvious that Eq.(\ref{Eq:imYZ0}) is always positive.

\section{Singularity conditions of pure mass}
\label{app:mass pure}

In this appendix, we aim to obtain the singularity conditions expressed only in terms of masses.
Here we introduce two angle variables to replace the two invariant mass variables used in the main text.
The first angle, named $\theta_{3/2}$, is the angle between the direction of particle 3 in the rest frame of particle 2 and the direction of particle 2 in the rest frame of the initial particle. 
The other angle is called $\theta_{4/1}$, which is the angle between the direction of particle 4 in the rest frame of particle 1 and the direction of particle 1 in the rest frame of the initial particle. 
The advantage of choosing these two angles is that they are completely independent and both range from 0 to $\pi$.

Firstly, we consider the case of Fig.~\ref{fg:3loopcuta}.
Our task is to obtain Eqs.(\ref{eq:invariantmassa}) and (\ref{eq:invariantmassb}) represented by these two new variables $\theta_{f_1/ 2}$ and $\theta_{4/ 1}$.
Here, we list the variables involved as:
\begin{align}
E_1 =& \frac{M^2+m_1^2-m_2^2}{2M}, \label{eq:E1}\\
E_2 =& \frac{M^2+m_2^2-m_1^2}{2M}, \label{eq:E2}\\
v_{1} =& \frac{\zeta(M,m_{1},m_{2})}{M^2-m^2_{2}+m^2_{1}}, \label{eq:v1}\\
v_{2} =& \frac{\zeta(M,m_{1},m_{2})}{M^2-m^2_{1}+m^2_{2}}, \label{eq:v2}\\
v_{3/ 2} =& \frac{\zeta(m_2,m_{3},m_{f_1})}{m^2_2-m^2_{f_1}+m^2_{3}}, \label{eq:v32}\\
v_{4/ 1} =& \frac{\zeta(m_{f_3},m_{4},m_{1})}{|m^2_1-m^2_{f_3}+m^2_{4}|}. \label{eq:v41}
\end{align}
%
%
Then we can obtain the velocity and energy of particles 3 and 4 in the rest frame of the initial particle with $\theta_{3/2}$ and $\theta_{4/1}$ as variables,
\begin{align}
v_{3\perp 1} =& \frac{v_{3/2}\sin\theta_{3/2}\sqrt{1-v_2^2}}{1+v_{3/2}v_2\cos\theta_{3/2}}, \label{eq:v3perp1}\\
v_{3\parallel 1} =& \frac{-v_{3/2}\cos\theta_{3/2}-v_2}{1+v_{3/2}v_2\cos\theta_{3/2}}, \label{eq:v3para1}\\
E_3 =& \frac{m_3(1+v_{3/2}v_2\cos\theta_{3/2})}{\sqrt{1-v_2^2}\sqrt{1-v^2_{3/2}}}, \label{eq:E3}\\
v_{4\perp 1} =& -\frac{v_{4/1}\sin\theta_{4/1}\sqrt{1-v_1^2}}{1+v_{4/1}v_1\cos\theta_{4/1}}, \label{eq:v4perp1}\\
v_{4\parallel 1} =& \frac{v_{4/1}\cos\theta_{4/1}+v_1}{1+v_{4/1}v_1\cos\theta_{4/1}}, \label{eq:v4para1}\\
E_4 =& \frac{m_4(1+v_{4/1}v_1\cos\theta_{4/1})}{\sqrt{1-v^2_1}\sqrt{1-v^2_{4/1}}}. \label{eq:E4}
\end{align}
Note that $v_{3\perp 1}>0>v_{4\perp 1}$ is valid here, which is part of Eq.(\ref{eq:conditionlast}), but the signs of $v_{3\parallel 1}$ and $v_{4\parallel 1}$ are determined by the variables $\theta_{3/2}$ and $\theta_{4/1}$ respectively.
In practice, we take the direction of particle 1 as the positive $\hat z$-axis. 
Since all particles are in the same plane, only two components are needed for each velocity.
Eq.(\ref{eq:invariantmassa}) can be written as:
\begin{align}
m^2_{f_2} =& (E_3-E_4)^2 -(E_3 v_{3\parallel 1}-E_4 v_{4\parallel 1})^2 \nonumber \\
& -(E_3 v_{3\perp 1}-E_4 v_{4\perp 1})^2.
\label{eq:invariantmassbatheta}
\end{align}
So we have:
\begin{align}
&\frac{m_3^2+m_4^2-m^2_{f_2}}{2m_3m_4}\sqrt{(1-v_2^2)(1-v_1^2)(1-v_{3/2}^2)(1-v_{4/1}^2)}\nonumber\\
=&(1+v_1v_2)+(v_1+v_2)v_{3/2}\cos\theta_{3/2}\nonumber \\
&+(v_1+v_2)v_{4/1}\cos\theta_{4/1}\nonumber\\
&+(1+v_1v_2)v_{3/2}v_{4/1}\cos\theta_{3/2}\cos\theta_{4/1}\nonumber\\
&+v_{3/2}v_{4/1}\sqrt{1-v_2^2}\sqrt{1-v_1^2}\sin\theta_{3/2}\sin\theta_{4/1}, 
\label{eq:condition222}
\end{align}
On the other hand, Eq.(\ref{eq:invariantmassb}) becomes:
\begin{align}
0 <& -\frac{v_1+v_2}{v_{3/2}}\sin\theta_{4/1}- (1+v_1v_2)\cos\theta_{3/2}\sin\theta_{4/1} \nonumber \\
& +\sqrt{1-v_1^2}\sqrt{1-v_2^2}\sin\theta_{3/2}\cos\theta_{4/1}.
\label{eq:condition111}
\end{align}
Thus, the relationship between the two angles is
\begin{align}
0\le\theta_{4/1}<\text{arccot}\left\{\frac{v_1+v_2+v_{3/2}(1+v_1v_2)\cos\theta_{3/2}}{v_{3/2}\sqrt{1-v_1^2}\sqrt{1-v_2^2}\sin\theta_{3/2}}\right\}.
\label{eq:condition112}
\end{align}
To obtain the mass constraints, it is necessary to determine whether there are $\theta_{3/2}$ and $\theta_{4/1}$ that satisfy Eq.(\ref{eq:condition112}) for a given set of masses so that Eq.(\ref{eq:condition222}) holds.
Since the left part of Eq.(\ref{eq:condition222}) is independent of the two angles, the problem is equivalent to finding the maximum and minimum values of the right side of Eq.(\ref{eq:condition222}) with the two angles as variables and requiring the left side to be between these maximum and minimum values, thus giving a new inequality.
This inequality provides the mass relationship that satisfies the aforementioned conditions.
For simplicity, we define:
\begin{align}
& A^{\pm}(\theta_{3/2}, \theta_{4/1})\nonumber\\
\equiv&(1+v_1v_2)+(v_1+v_2)v_{3/2}\cos\theta_{3/2}
\nonumber\\
&+(v_1+v_2)v_{4/1}\cos\theta_{4/1}\nonumber\\
&+(1+v_1v_2)v_{3/2}v_{4/1}\cos\theta_{3/2}\cos\theta_{4/1}\nonumber\\
&\pm v_{3/2}v_{4/1}\sqrt{1-v_2^2}\sqrt{1-v_1^2}\sin\theta_{3/2}\sin\theta_{4/1}. 
\label{eq:Apm}
\end{align}
Then, our goal is to find the maximum and minimum values of $A^+(\theta_{3/2}, \theta_{4/1})$.

It can be found that the derivative of $A^+(\theta_{3/2}, \theta_{4/1})$ with respect to variable $\theta_{4/1}$ is:
\begin{align}
& \frac{1}{v_{4/1}v_{3/2}}\frac{\partial A^+(\theta_{3/2}, \theta_{4/1})}{\partial \theta_{4/1}} \nonumber \\
=& -\frac{v_1+v_2}{v_{3/2}}\sin\theta_{4/1}-(1+v_1v_2)\cos\theta_{3/2}\sin\theta_{4/1} \nonumber \\
& +\sqrt{1-v_2^2}\sqrt{1-v_1^2}\sin\theta_{3/2}\cos\theta_{4/1}, 
\label{eq:condition2221}
\end{align}
which is exactly the right side of Eq.(\ref{eq:condition111}), so it is required to be positive.
In other word, when $\theta_{4/1}$ satisfies Eq.(\ref{eq:condition112}), $A^+(\theta_{3/2}, \theta_{4/1})$ is monotonically increasing with respect to $\theta_{4/1}$.
For a given $\theta_{3/2}$, the minimum value of $A^+(\theta_{3/2}, \theta_{4/1})$ is at $\theta_{4/1}=0$, i.e.,
\begin{align}
A^+(\theta_{3/2}, 0)=& (1+v_1v_2)+(v_1+v_2)v_{3/2}\cos\theta_{3/2} \nonumber \\
&+ (v_1+v_2)v_{4/1} \nonumber \\
&+ (1+v_1v_2)v_{3/2}v_{4/1}\cos\theta_{3/2}.
\label{eq:Aplus}
\end{align}
Obviously when $\theta_{3/2}=\pi$, Eq.(\ref{eq:Aplus}) reaches its minimum value, which is:
\begin{align}
A^+_{\text{min}}\equiv A(\pi, 0)=& (1+v_1v_2)(1-v_{3/2}v_{4/1}) \nonumber \\
&+(v_1+v_2)(v_{4/1}-v_{3/2}).
\end{align}
The maximum value of $A^+(\theta_{3/2}, \theta_{4/1})$ is clearly at $\theta_{3/2}=0$ and $\theta_{4/1}=0$, because all the coefficients before the cosine function in Eq.(\ref{eq:condition222}) are positive, and the coefficient of the product of two cosine functions is always greater than the coefficient of the product of two sine functions.
We obtain,
\begin{align}
A^+_{\text{max}}\equiv A(0, 0) =& (1+v_1v_2)(1+v_{3/2}v_{4/1}) \nonumber \\
&+(v_1+v_2)(v_{4/1}+v_{3/2}).
\end{align}

The left side of Eq.(\ref{eq:condition222}) should be in the range $[A^+_{\text{min}}, A^+_{\text{max}}]$, which leads to
\begin{align}
&F^{-}\le 4m_1^2m_2^2(m_3^2+m_4^2-m_{f_2}^2)\le F^{+},
\end{align}
where
\begin{align}
F^{\pm}\equiv&(M^2-m_1^2-m_2^2) \times\nonumber\\
&\,\,\,\left[(m_2^2-m_{f_1}^2+m_3^2)|m_1^2-m_{f_3}^2+m_4^2|\right. \nonumber\\
&\,\,\,\,\,\,\left.
\pm\zeta(m_{f_3},m_4,m_1)\zeta(m_2,m_3,m_{f_1}) \right]\nonumber\\
&+\zeta(M,m_1,m_2)\times\nonumber\\
&\,\,\,\left[(m_2^2-m_{f_1}^2+m_3^2)\zeta(m_{f_3},m_4,m_1)\right. \nonumber\\
&\,\,\,\,\,\,\left.\pm|m_1^2-m_{f_3}^2+m_4^2|\zeta(m_2,m_3,m_{f_1})\right].
\end{align}
This is the singularity condition of Fig.~\ref{fg:3loopa} and note that the on-shell conditions are always required.

Next we turn to the mass condition for the case of Fig.~\ref{fg:3loopb}.
As discussed before, there is a difference at this point where $v_{4\perp1}>0$ is required, and of course $v_{3\perp1}$ is always positive.
So $v_{4\perp1}$ becomes
\begin{align}
\tilde{v}_{4\perp 1}= \frac{v_{4/1}\sin\theta_{4/1}\sqrt{1-v_1^2}}{1+v_{4/1}v_1\cos\theta_{4/1}}.
\label{eq:v4perp1b}
\end{align}
Expect for Eq.(\ref{eq:v4perp1}), the other energies and velocities in Eqs.(\ref{eq:E1}-\ref{eq:E4}) are still applicable.
Eq.(\ref{eq:invariantmassbatheta}) becomes
\begin{align}
m^2_{f_2}=& (E_3+E_4)^2-(E_3 v_{3\parallel 1}+E_4 v_{4\parallel 1})^2 \nonumber \\
& -(E_3 v_{3\perp 1}+E_4 \tilde{v}_{4\perp 1})^2,
\label{eq:invariantmassbathetb}
\end{align}
which leads to 
\begin{align}
&\frac{m^2_{f_2}-m_3^2-m_4^2}{2m_3m_4}\sqrt{(1-v_2^2)(1-v_1^2)(1-v_{3/2}^2)(1-v_{4/1}^2)} \nonumber \\
=& A^-(\theta_{3/2}, \theta_{4/1}).
\label{eq:condition222b}
\end{align}

We will then consider the conditions under which singularities occur, as given in Eqs.(\ref{eq:invariantmassbb}, \ref{eq:invariantmassbb2}) or Eqs.(\ref{eq:v4gv3111}, \ref{Eq:patternb111}, \ref{eq:v4gv3222}, \ref{Eq:patternb222b}).
As discussed at the beginning of Sec.~\ref{sec:diagramB}, there are two cases here that can be correlated by exchanging particle indices.
We start with Eqs.(\ref{eq:v4gv3111}, \ref{Eq:patternb111}) to derive the relationship between the masses, and then we can naturally obtain the formulae for the other case.
Eqs.(\ref{eq:v4gv3111}, \ref{Eq:patternb111}) are:
\begin{align}
& \frac{v_{4/1}\sin\theta_{4/1}\sqrt{1-v_1^2}}{1+v_{4/1}v_1\cos\theta_{4/1}}> \frac{v_{3/2}\sin\theta_{3/2}\sqrt{1-v_2^2}}{1+v_{3/2}v_2\cos\theta_{3/2}}, \label{eq:v41v31b}
\end{align}
\begin{align}
0>& \frac{v_1+v_2}{v_{3/2}}\sin\theta_{4/1}+(1+v_1v_2)\cos\theta_{3/2}\sin\theta_{4/1} \nonumber \\
& +\sqrt{1-v_1^2}\sqrt{1-v_2^2}\sin\theta_{3/2}\cos\theta_{4/1}.
\label{eq:cb222b}
\end{align}
We also need to find the minimum and maximum values of $A^-(\theta_{3/2}, \theta_{4/1})$ when $\theta_{3/2}$ and $\theta_{4/1}$ satisfy Eqs.(\ref{eq:v41v31b}) and (\ref{eq:cb222b}), so that we can obtain the pure mass condition.
Compared to the previous situation, there is an additional restriction from Eq.(\ref{eq:v41v31b}) for these two angles. 
We first ignore this restriction and try to obtain the maximum and minimum values of $A^-(\theta_{3/2}, \theta_{4/1})$, and then check whether the angles satisfy Eq.(\ref{eq:v41v31b}) at this point.
Similar to Eq.(\ref{eq:condition2221}), we have:
\begin{align}
&\frac{1}{v_{4/1}v_{3/2}}\frac{\partial A^-(\theta_{3/2}, \theta_{4/1})}{\partial \theta_{4/1}} \nonumber \\
=& -\frac{v_1+v_2}{v_{3/2}}\sin\theta_{4/1}-(1+v_1v_2)\cos\theta_{3/2}\sin\theta_{4/1} \nonumber \\
& -\sqrt{1-v_2^2}\sqrt{1-v_1^2}\sin\theta_{3/2}\cos\theta_{4/1}, 
\label{eq:condition2221b}
\end{align}
and by comparing it with Eq.(\ref{eq:cb222b}), we can see that it is always positive.
Thus, there are:
\begin{align}
&\pi\ge\theta_{4/1}>\theta_{\text{min}}, \nonumber \\
&\theta_{\text{min}}\equiv\text{arccot}\left\{-\frac{\frac{v_1+v_2}{v_{3/2}}+(1+v_1v_2)\cos\theta_{3/2}}{\sqrt{1-v_1^2}\sqrt{1-v_2^2}\sin\theta_{3/2}}\right\}.
\label{eq:condition112b}
\end{align}
For a given $\theta_{3/2}$, the maximum value of $A^-(\theta_{3/2}, \theta_{4/1})$ is taken when $\theta_{4/1}=\pi$, that is,
\begin{align}
A^{-}(\theta_{3/2}, \pi) &= (1+v_1v_2)+(v_1+v_2)v_{3/2}\cos\theta_{3/2} \nonumber \\
&-(v_1+v_2)v_{4/1}-(1+v_1v_2)v_{3/2}v_{4/1}\cos\theta_{3/2}, 
\end{align}
Then it can be found that the maximum value of $A^-(\theta_{3/2}, \pi)$ is at $\theta_{3/2}=0$ or $\pi$, i.e.,
\begin{align}
A^-_{\text{max}}\equiv&\left\{
\arraycolsep=4pt\def\arraystretch{1.3}
\begin{array}{ll}
A^-(0, \pi),&  \tilde{v}\equiv\frac{v_1+v_2}{1+v_1v_2}>v_{4/1} \\
A^-(\pi, \pi),&  \tilde{v}\equiv\frac{v_1+v_2}{1+v_1v_2}<v_{4/1}  
\end{array}
\right..\label{eq:Ammax}
\end{align}
Here, $\theta_{3/2}=0$ or $\pi$ and $\theta_{4/1}=\pi$ are exactly at the boundaries of Eq.(\ref{eq:v41v31b}), so $A^-_{\text{max}}$ is indeed the maximum value when the two angles satisfy Eqs.(\ref{eq:v41v31b}) and (\ref{eq:condition112b}).

For the minimum value of $A^-$, $\theta_{4/1}$ takes the minimum value $\theta_{\text{min}}$ defined in Eq.(\ref{eq:condition112b}), so we have:
\begin{align}
&A^{-}(t)\equiv A^{-}(\theta_{3/2}, \theta_{\text{min}}) \nonumber \\
=&v_{3/2}t -v_{4/1}v_{3/2}\sqrt{t^2+(1-v_1^2)(1-v_2^2)(1-\frac{1}{v^2_{3/2}})}, \\
&t(\cos\theta_{3/2})\equiv \frac{(1+v_1v_2)}{v_{3/2}}+(v_1+v_2)\cos\theta_{3/2}.
\end{align}
Here $t$ is always positive and in the range of $[t(-1), t(1)]$.
If we do not consider the range of $t$ for now, then the minimum value of $A^{-}$ is at $t=\tilde{t}$, where
\begin{align}
\tilde{t}\equiv&\sqrt{\frac{(1-v_1^2)(1-v_2^2)(1-v_{3/2}^2)}{v_{3/2}^2(1-v_{4/1}^2)}}.
\end{align}
This is solved from ${\rm d}A^{-}(t)/{\rm d}t = 0$.
When $\tilde{t}$ is outside the range $[t(-1), t(1)]$, the minimum of $A^-$ should be at $\theta_{3/2}=\pi$ or $0$.
It should be noted that from Eq.(\ref{eq:condition112b}), $\theta_{\text{min}}$ can be $0$ or $\pi$ depending on the sign of $(v_1+v_2)/(1+v_1v_2)-v_{3/2}$ for $\theta_{3/2}=\pi$, while $\theta_{\text{min}}$ can only take $\pi$ for $\theta_{3/2}=0$.
In summary, there are four possible cases for the minimum value of $A^-$, which are:
\begin{align}
A^-_{\text{min}}\equiv&\left\{
\arraycolsep=4pt\def\arraystretch{1.3}
\begin{array}{ll}
A^-(0, \pi), &  \tilde{t}>t(1) \\
A^-(\tilde{t}),& t(-1) \le\tilde{t}\le t(1) \\
A^-(\pi, 0),&  \tilde{t}<t(-1)\,\&\, \tilde{v}<v_{3/2} \\
A^-(\pi, \pi),&  \tilde{t}<t(-1)\,\&\,  \tilde{v}> v_{3/2}
\end{array}
\right..
\end{align}
Except for the case of $t=\tilde{t}$, all other cases are $\theta_{3/2}, \theta_{4/1}=\pi$ or $0$, which are taken at the boundary of Eq.(\ref{eq:v41v31b}), ensuring that this condition is satisfied in these cases.
For $t=\tilde{t}$, the minimum value of $A^-$ is $A^-(\tilde{t})$. 
%
%
In fact, it can be seen that the condition $t(-1) \le\tilde{t}\le t(1)$ remains the same after exchanging particles 1 and 2, 3 and 4, $f_1$ and $f_3$.
%
%
We find that if $m_{f_2}>m_3+m_4$, the left part of Eq.(\ref{eq:condition222b}) is always greater than $A^-(\tilde{t})$, so we can directly get this minimum value as 0.

We consider $A^-_{\text{max}}$ and $A^-_{\text{min}}$ together and compare the conditions in Eq.(\ref{eq:Ammax}) to find that 
\begin{align}
\tilde{t}>t(1)\,\,&\to\,\, v_{4/1}>\frac{\tilde{v}+v_{3/2}}{1+\tilde{v}v_{3/2}}>\tilde{v},\\
\tilde{t}<t(-1)\,\,\&\,\, \tilde{v}>v_{3/2}\,\,&\to\,\,  
v_{4/1}<\frac{\tilde{v}-v_{3/2}}{1-\tilde{v}v_{3/2}}<\tilde{v}.
\end{align}
Obviously, the maximum and minimum values of $A^-$ cannot be the same.
Therefore, combining Eq.(\ref{eq:condition222b}), we have:
\begin{align}
&\frac{m^2_{f_2}-m_3^2-m_4^2}{2m_3m_4}\sqrt{(1-v_2^2)(1-v_1^2)(1-v_{3/2}^2)(1-v_{4/1}^2)} \nonumber\\
\in& \left\{
\arraycolsep=4pt\def\arraystretch{1.3}
\begin{array}{l}
(A^-(0, \pi),A^-(\pi, \pi)),\,\,   v_{4/1}>\frac{\tilde{v}+v_{3/2}}{1+\tilde{v}v_{3/2}}\\
(0,A^-(\pi, \pi)),\,\, \text{min}(\frac{|\tilde{v}-v_{3/2}|}{1-\tilde{v}v_{3/2}},\tilde{v}) < v_{4/1}\le\frac{\tilde{v}+v_{3/2}}{1+\tilde{v}v_{3/2}} \\
(0,A^-(0, \pi)),\,\,  \frac{|\tilde{v}-v_{3/2}|}{1-\tilde{v}v_{3/2}} < v_{4/1}\le\tilde{v} \\
(A^-(\pi, 0),A^-(\pi, \pi)),\,\, \tilde{v}<v_{4/1}<\frac{-\tilde{v}+v_{3/2}}{1-\tilde{v}v_{3/2}} \\
(A^-(\pi, 0),A^-(0, \pi)),\,\, v_{4/1}<\text{min}(\tilde{v},\,\frac{-\tilde{v}+v_{3/2}}{1-\tilde{v}v_{3/2}}) \\
( A^-(\pi, \pi),A^-(0, \pi)),\,\,  v_{4/1}\le\frac{\tilde{v}-v_{3/2}}{1-\tilde{v}v_{3/2}}   
\end{array}
\right..
\end{align}
Then the mass condition of singularity can be straightforwardly obtained as:
\begin{align}
&4m_1^2m_2^2(m^2_{f_2}-m_3^2-m_4^2)\nonumber\\
\in& \left\{
\arraycolsep=4pt\def\arraystretch{1.3}
\begin{array}{l}
(\tilde{F}^-,\tilde{F}^+),\,\,   v_{4/1}>\frac{\tilde{v}+v_{3/2}}{1+\tilde{v}v_{3/2}}\\
(0,\tilde{F}^+),\,\, \text{min}(\frac{|\tilde{v}-v_{3/2}|}{1-\tilde{v}v_{3/2}},\tilde{v}) < v_{4/1}\le\frac{\tilde{v}+v_{3/2}}{1+\tilde{v}v_{3/2}} \\
(0,\tilde{F}^-),\,\,  \frac{|\tilde{v}-v_{3/2}|}{1-\tilde{v}v_{3/2}} < v_{4/1}\le\tilde{v} \\
(F^-,\,\tilde{F}^+),\,\, \tilde{v}<v_{4/1}<\frac{-\tilde{v}+v_{3/2}}{1-\tilde{v}v_{3/2}} \\
(F^-,\tilde{F}^-),\,\, v_{4/1}<\text{min}(\tilde{v},\,\frac{-\tilde{v}+v_{3/2}}{1-\tilde{v}v_{3/2}}) \\
( \tilde{F}^+,\tilde{F}^-),\,\,  v_{4/1}\le\frac{\tilde{v}-v_{3/2}}{1-\tilde{v}v_{3/2}}   
\end{array}
\right..
\end{align}
where,
\begin{align}
\tilde{F}^{\pm}\equiv&\left[(m_2^2-m_{f_1}^2+m_3^2)|m_1^2-m_{f_3}^2+m_4^2|\right. \nonumber\\
&\left.\pm\zeta(m_{f_3},m_4,m_1)\zeta(m_2,m_3,m_{f_1})\right](M^2-m_1^2-m_2^2) \nonumber\\
&-\zeta(M,m_1,m_2)\left[(m_2^2-m_{f_1}^2+m_3^2)\zeta(m_{f_3},m_4,m_1)\right. \nonumber\\
&\left.\pm|m_1^2-m_{f_3}^2+m_4^2|\zeta(m_2,m_3,m_{f_1})\right].
\end{align}
At last, for $\tilde{I}_{b_2}$, as mentioned before, we just need to swap the particle indices in the equations.
These are all for Fig.~\ref{fg:3loopb}, and it should be noted again that the on-shell conditions are always required.

\end{appendix}
\clearpage

\bibliographystyle{plain}

\end{document}